\documentclass[aps,prl,twocolumn,superscriptaddress,showpacs,amsmath,amssymb,floatfix]{revtex4}

\usepackage[english]{babel}
\usepackage{amsmath}
\usepackage{amssymb}
\usepackage{graphicx,color}
\usepackage{epsfig}

\newcommand{\Ham}{\mathcal H}
\newcommand{\ket}[1]{\left\vert#1\right\rangle}
\newcommand{\bra}[1]{\left\langle#1\right\vert}
\newcommand{\beq}{\begin{equation}}
\newcommand{\eeq}{\end{equation}}

\begin{document}

\title{Quantum quenches, linear response and superfluidity out of equilibrium}

\author{Davide Rossini}
\affiliation{NEST, Scuola Normale Superiore \& Istituto Nanoscienze-CNR, I-56126 Pisa, Italy} 

\author{Rosario Fazio}
\affiliation{NEST, Scuola Normale Superiore \& Istituto Nanoscienze-CNR, I-56126 Pisa, Italy} 

\author{Vittorio Giovannetti}
\affiliation{NEST, Scuola Normale Superiore \& Istituto Nanoscienze-CNR, I-56126 Pisa, Italy} 

\author{Alessandro Silva}
\affiliation{Scuola Internazionale Superiore di Studi Avanzati (SISSA), I-34136 Trieste, Italy}
\affiliation{International Centre for Theoretical Physics (ICTP), I-34151 Trieste, Italy}


\begin{abstract}
  By analysing the sensitivity to a twist in the boundary conditions of the stationary state 
  attained by a many-body system long after a quantum quench, we extend the concepts 
  of the helicity modulus and the stiffness to non-equilibrium situations.
  Using these generalised quantities, we characterise the out-of-equilibrium dynamics 
  of hard-core bosons quenched to/from superfluid/insulating phases and show that 
  qualitative new features emerge as compared to the equilibrium case. 
  Our predictions can be tested in experiments with cold bosonic atoms confined in 
  toroidal traps and subject to artificial gauge fields. 
\end{abstract}

\maketitle

{\bf Introduction. --- }
The impressive experimental advances in cold atomic 
physics have given the opportunity to study non-equilibrium dynamics 
of closed many-body systems with an unprecedented degree of control 
and tunability~\cite{Bloch_2008, Polkovnikov_2011}. 
The lack of ergodicity in two colliding clouds~\cite{Kinoshita_2006}, 
the observation of collapse and revivals in a system driven across the 
Mott insulator-to-superfluid phase transition~\cite{Greiner_2002}, 
the prethermalisation of one-dimensional condensates~\cite{Gring_2012}, 
the light-cone spreading of correlations~\cite{Cheneau_2012} 
are only few important examples of the ongoing experimental activity in this area.

One of the most important problems in this field is the characterisation 
of the stationary state attained by the system after a quantum quench. 
Intuition suggests that even an isolated many-body system 
should be able to reach, at long times, a thermal steady state compatible 
with the initial energy density, as far as local observables 
are concerned~\cite{Deutsch_1991,Sredincki_1994,Rigol_2008}. 
This expectation is justified by the fact that, typically, a large 
system should act as its own environment. Notable exceptions are integrable 
systems~\cite{Calabrese_2006,Barthel_2008,Rigol_2007,Cramer_2010,Calabrese_2012}, 
which appear to relax to a steady state keeping track of all 
the constants of motion: the generalised Gibbs ensemble (GGE)~\cite{Rigol_2007}. 
Differences between thermal and GGE states are frequently only 
quantitative~\cite{Rossini_2009,Kennes_2010,Fagotti2011,Mussardo2013}. 
Finding ways to neatly detect signatures of the GGE is highly desirable. 
Furthermore, since in almost integrable systems 
full thermalisation is preceded by a metastable non-equilibrium 
prethermalised state~\cite{Berges2004, Gring_2012} closely related 
to the GGE~\cite{Kollar2011}, an unambiguous identification of non-thermal 
behaviour is important also to study this type of two-stage dynamics.

While so far expectation values of observables in the stationary state 
have been studied, in (quasi-)equilibrium systems the response 
to external (static or dynamic) perturbations can be as important to identify
the system properties~\cite{Foini_2011, Essler_2012, Russomanno_2013, Marcuzzi_2013}. 
Here we propose a characterisation of stationary states at long times 
after a quench in terms of time averaged response functions
that measure the sensitivity to a twist in the boundary conditions, thus extending 
the concepts of stiffness (Drude peak)~\cite{Kohn_1964} and helicity 
modulus~\cite{Fisher_1973} to non-equilibirum conditions. 
This approach becomes particularly interesting in low dimensions, 
where it provides a powerful tool to characterise superfluid correlations 
(see also Ref.~\cite{Eggel_2011}).

Both the out-of-equilibrium stiffness and the helicity modulus are introduced 
in an operational way that can, in principle, be implemented experimentally. 
By studying in detail these quantities in two integrable models 
of hard-core bosons, we show that non-thermal steady states 
attained after a quench have properties that depart quite dramatically 
from what is predicted in equilibrium. The stiffness can become negative. 
The helicity modulus may remain finite even in the insulating state. 
This behaviour is in striking contrast to what is expected for non-integrable systems, 
where thermalisation occurs~\cite{note_3}. 
At the end of the paper we discuss the possibility to measure both quantities experimentally.

{\bf Out-of-equilibrium linear response. --- }
Before introducing 
the stiffness and the helicity modulus in a non-equilibrium setting, it is necessary 
to discuss linear response for quantum systems out of equilibrium.
Linear response to external perturbations is a primary tool to address 
the equilibrium properties of many-body systems. 
Usually one considers the response to a time-dependent 
perturbation $\hat{V}(t)$ of an otherwise stationary observable 
$\langle \hat{A}(t) \rangle = \langle \hat{A} \rangle_{\rm eq} +\langle \delta \hat{A}(t)\rangle$.
This can be easily generalised to situations where 
the dynamics is non-trivial even without perturbation. 
Of interest for us is the following case: after initialising the system 
in the zero-temperature ground state of Hamiltonian $\hat{\Ham}(\lambda_i)$, 
at $t=0$ we perform a {\it sudden quench} 
$\lambda_i \rightarrow \lambda_f$, simultaneously switching on 
a weak time-dependent ({\it e.g.}, oscillatory) perturbation $\hat{V}(t)$. 
In this setup it is natural to study the time-dependent linear response 
of the system on top of the non-equilibrium dynamics resulting from the quench. 
A sketch of this protocol is illustrated in~\cite{SI}.

As in the equilibrium case, the key quantity, 
given an observable $\hat{A}$ and a perturbation $\hat{V}(t)=h(t) \hat{B}$, 
is~\cite{SI} the response function 
$\chi_{AB}=-i\theta(t-t') \langle [\hat{A}(t),\hat{B}(t')] \rangle_0$. 
All operators are evolved with the final Hamiltonian $\hat{\Ham}(\lambda_f)$ 
and $\langle \cdot \rangle_0= \bra{\psi_0(\lambda_i)} \cdot \ket{\psi_0(\lambda_i)}$, 
$\ket{\psi_0(\lambda_i)}$ being the ground state of the initial Hamiltonian.
In usual linear response theory $\chi_{AB}$ depends on time 
differences $t-t'$, making it possible to analyse the system properties, 
such as the presence of quasi-particles, by taking a Fourier transform 
and looking for features such as peaks or singularities. 
Here instead $\chi_{AB}$ depends on two separate times $t$ and $t'$. 
However, one may still effectively analyse the dynamics passing to Wigner coordinates 
$T=(t+t')/2$, $\tau=t-t'$ and Fourier transforming with respect to $\tau$,
restricted to $\vert \tau \vert < 2 T$ (we assume $t, t' > 0$). 
The resulting function $\chi_{AB}(T,\omega)$ may describe the formation, 
dynamics and relaxation of quasi-particles at frequencies $\omega$ much larger than the 
typical time scale associated to the quench dynamics. 
Moreover, from $\chi_{AB}(T,\tau)$ one can obtain information about the stationary 
state attained at long times. Indeed, by averaging over $T$, that is 
$\overline{[\, \cdot \,]}=\lim_{T\rightarrow+\infty} T^{-1} \int_0^T [\, \cdot \,]$, 
one gets~\cite{SI}:
\begin{equation}
  \chi_{\rm st}(\tau) \equiv \overline{\chi_{AB}} = -i \theta(\tau) 
  \Big\langle \! \big[ \hat{A} \! \left( \tau\right) \! , 
    \hat{B} \! \left( 0\right) \big] \! \Big\rangle_{\rm st}
\label{linear}
\end{equation} 
where $\langle \, \cdot \, \rangle_{\rm st} = {\rm Tr}[\hat{\rho}_{{\rm st}} \, \cdot \,]$, 
with $\hat{\rho}_{{\rm st}} \equiv \overline{\hat{\rho}(t)} = \sum_{n,r} p_n \, \hat{\pi}_n.$
The operator $\hat{\pi}_n = \sum_{r} \ket{\psi_{n,r}}\bra{\psi_{n,r}}$ projects
on the $n$-th eigenspace of $\hat{\Ham}(\lambda_f)$ (characterised by energy 
$\varepsilon_n$ and eigenstates $\ket{\psi_{n,r}}$, $r$ being a degeneracy index),
while the weights $p_n = \bra{\psi_0} \hat{\pi}_n \ket{\psi_0}$ are
the probabilities to find the initial state on that subspace.
The state $\hat{\rho}_{\rm st}$ is the so-called {\it diagonal ensemble}. 
The only formal difference between these response functions and the equilibrium ones
is the appearance of the diagonal ensemble instead of the Gibbs ensemble. 

Linear response, Eq.~\eqref{linear}, is needed to characterise superfluid 
correlations and transport properties of interacting bosons past a quench.
Superfluids, as well as ideal conductors, are characterised 
by their sensitivity to a twist in the boundary conditions. 
In equilibrium, the {\it helicity modulus} ${\cal Y}$~\cite{Fisher_1973}, 
{\it i.e.}, a thermodynamic quantity related to 
the second derivative of the free energy with respect to the twist, 
quantifies superfluid correlations and is particularly important 
in low dimensional systems, where the superfluid order parameter vanishes.
Sensitivity to phase twists is also important for normal quantum fluids, 
being related to the {\it charge stiffness} $\mathcal{D}$, 
{\it i.e.}, the amplitude of the Drude peak in electric conductivity~\cite{Kohn_1964}. 
The connection of stiffness and helicity modulus with integrability properties 
of the underlying models has been exploited in great detail
at equilibrium~\cite{Shastry_1990, Fye_1991, Giamarchi_1995, Castella_1995, Zotos_1999, Sirker_2011, Prosen_2013}. 
Here we extend their notion to non-equilibrium systems. 

From now on we consider a one-dimensional loop of length $L$ 
(the generalisation to $d$-dimensions is straightforward). 
The twist is induced by piercing the loop with an external flux. 
As we will see below, the stiffness and the helicity modulus are associated to the response 
of the system to an oscillating/static external magnetic field respectively~\cite{SI}.

{\bf Charge stiffness. --- }
In equilibrium the charge stiffness is related to the linear response of the system 
to a time-dependent magnetic flux. It quantifies the strength of the zero-frequency 
Drude weight in the conductivity~\cite{Kohn_1964, Castella_1995}. 
In the present case one should calculate the time averaged circulating current 
induced by an external flux after the stationary state has been attained. 
It is thus possible to access the {\it non-equilibrium stiffness}, 
or the amplitude of the Drude peak of the response function to a uniform time-dependent field 
in the stationary regime, using 
\begin{equation}
  {\cal D}_{\rm neq} \!=\! \frac{1}{L} \bigg\{ \!- \frac{\langle \hat{\cal T} \rangle}{2} - \! \sum_n \sum_{m \neq n \atop r,s} 
  p_n \frac{\vert \! \bra{\psi_{n,r}} \hat{\cal J}_{\varphi=0} \ket{\psi_{m,s}} \! \vert^2}{\varepsilon_m - \varepsilon_n} \bigg\} ,
  \label{nes}
\end{equation}
$\langle \hat{\cal T} \rangle$ being the expectation value 
of the kinetic-energy, $p_n$ the probabilities in the diagonal 
ensemble~\cite{note_0}, $\hat{\cal J}_\varphi = - \partial_\varphi \hat{\cal H}_\varphi$ 
the current operator, and $\varphi = \phi/L$ (with $\phi$ the flux piercing the loop 
or equivalently the twist in the boundary conditions~\cite{Fisher_1973}). 
Using second-order perturbation theory in the twist, it can be shown that 
the stiffness quantifies the curvature of energy levels in $\varphi$:
\begin{equation}
  {\cal D}_{\rm neq} = \frac{1}{2 L} \sum_n p_n \partial^2_\varphi \varepsilon_n(\varphi)_{\vert_{\varphi = 0}} \,.
  \label{eq:stiff1}
\end{equation}
This is the same expression as in equilibrium, provided that $p_n$ are replaced by Boltzmann weights.

{\bf Helicity modulus. --- }
The definition of helicity modulus ${\cal Y}$ requires 
a little bit of care. Indeed this is the coefficient connecting the current density $j$ 
to a static twist $\varphi$: 
$j {}_{\vert_{\varphi \to 0}} = -{\cal Y} \cdot \varphi$. 
In principle one can perform linear response as before. 
For our purposes it is however more transparent to proceed differently and consider 
the time averaged current density: 
$\overline{j_\varphi} = L^{-1} \overline{\bra{\psi(t)} {\cal \hat{J}_\varphi} \ket{\psi(t)}}$.
After expanding over the Hamiltonian eigenbasis, one finds
$\overline{j_\varphi} = - L^{-1} \sum_n p_n(\varphi) \, \partial_\varphi \varepsilon_n(\varphi)$. 
It is then natural to define the {\it non-equilibrium helicity modulus}: 
\begin{equation}
  {\cal Y}_{\rm neq} = -\partial_{\varphi} \overline{j_\varphi}_{\vert_{\varphi = 0}} \,.
  \label{neh}
\end{equation}
This definition reduces to the one at equilibrium, if time-averaged 
probabilities $p_n(\varphi)$ are again replaced by Boltzmann weights. 
However ${\cal Y}_{\rm neq}$ {\em cannot} be obtained from the equilibrium 
helicity modulus (the explicit expression for ${\cal Y}_{\rm eq}$
is given in~\cite{note_1}) by replacing Boltzmann weights with $p_n(\varphi)$. 
The protocol we defined above leading to Eq.~\eqref{neh} is the appropriate way to do define it. 

Eqs.~\eqref{eq:stiff1} and~\eqref{neh} quantify the phase rigidity of superfluid 
and normal systems following a quench. In equilibrium at zero temperature 
and for time-reversal invariant systems ${\cal D}_{\rm eq} = {\cal Y}_{\rm eq}$, 
moreover in thermal equilibrium ${\cal Y}_{\rm eq}, {\cal D}_{\rm eq} \geq 0$. 
These properties {\em no longer hold} for ${\cal Y}_{\rm neq}$ 
and ${\cal D}_{\rm neq}$~\cite{note_1}. We expect these differences to appear 
in integrable systems. On the contrary they should not be relevant when 
thermalisation occurs. 

{\bf Examples. --- }
We now discuss the behaviour of ${\cal Y}_{\rm neq}$ and ${\cal D}_{\rm neq}$
in two relevant cases of bosonic systems. In both examples the underlying Hamiltonian 
is integrable and describes tight-binding bosons hopping in a one-dimensional ring 
with $L$ sites, subject to an external flux: 
\begin{equation}
  \hat{\cal H}(\varphi) = \sum_j \; ( e^{i\varphi} \hat{b}^{\dagger}_{j+1} \hat{b}_j + {\rm H.c.} ) 
  + \hat{\cal V}_{\rm int} \;.
  \label{eq:Ham}
\end{equation}
Here $\hat{b}^\dagger_j \, (\hat{b}_j)$ are hard-core bosonic creation 
(annihilation) operators on the $j$-th lattice site. 
The first term in $\hat{\cal H}(\varphi)$ is the kinetic energy, 
while $\hat{\cal V}_{\rm int}$ is an additional contribution specified below.
We study: $i)$ hard-core bosons in a staggered field 
$\hat{\cal V}_{\rm int}^{\rm (I)} = V \sum_j (-1)^j \hat{n}_j $;
$\; ii)$ interacting bosons 
with $\hat{\cal V}_{\rm int}^{\rm (II)} = V \sum_j \hat{n}_j \hat{n}_{j+1}$,
where $\hat{n}_j = \hat{b}^\dagger_j \hat{b}_j$ is the boson number operator.
We consider half filling and perform a quench in interaction strength, 
from an initial value $V_i$ to a final $V_f$.

Model I is equivalent to a spin-$1/2$ XX chain 
in a staggered transverse magnetic field, and can be diagonalised through 
a Jordan Wigner transformation~\cite{Lieb_1961}. 
In its diagonal form (in momentum space) the Hamiltonian reads:
$\hat{\Ham} = \sum_{\mid k \mid< \pi/2} \epsilon_k \big( \hat{\gamma}_k^+ {}^\dagger \, \hat{\gamma}^+_k 
- \hat{\gamma}^-_k {}^\dagger \, \hat{\gamma}^-_k \big)$, 
where $\hat{\gamma}^\pm_k$ are fermionic quasiparticle operators, 
$\epsilon_k=\sqrt{4 \cos ( k-\varphi )^2 + V^2}$. 
At half filling its ground-state phase diagram presents a gapped insulating phase 
for all $V \neq 0$, separated by a superfluid point at $V=0$. 
Being easily solvable, it allows us to catch all the salient features 
which we will also find later by solving numerically model II. 

\begin{figure}
  {\includegraphics[width=\columnwidth]{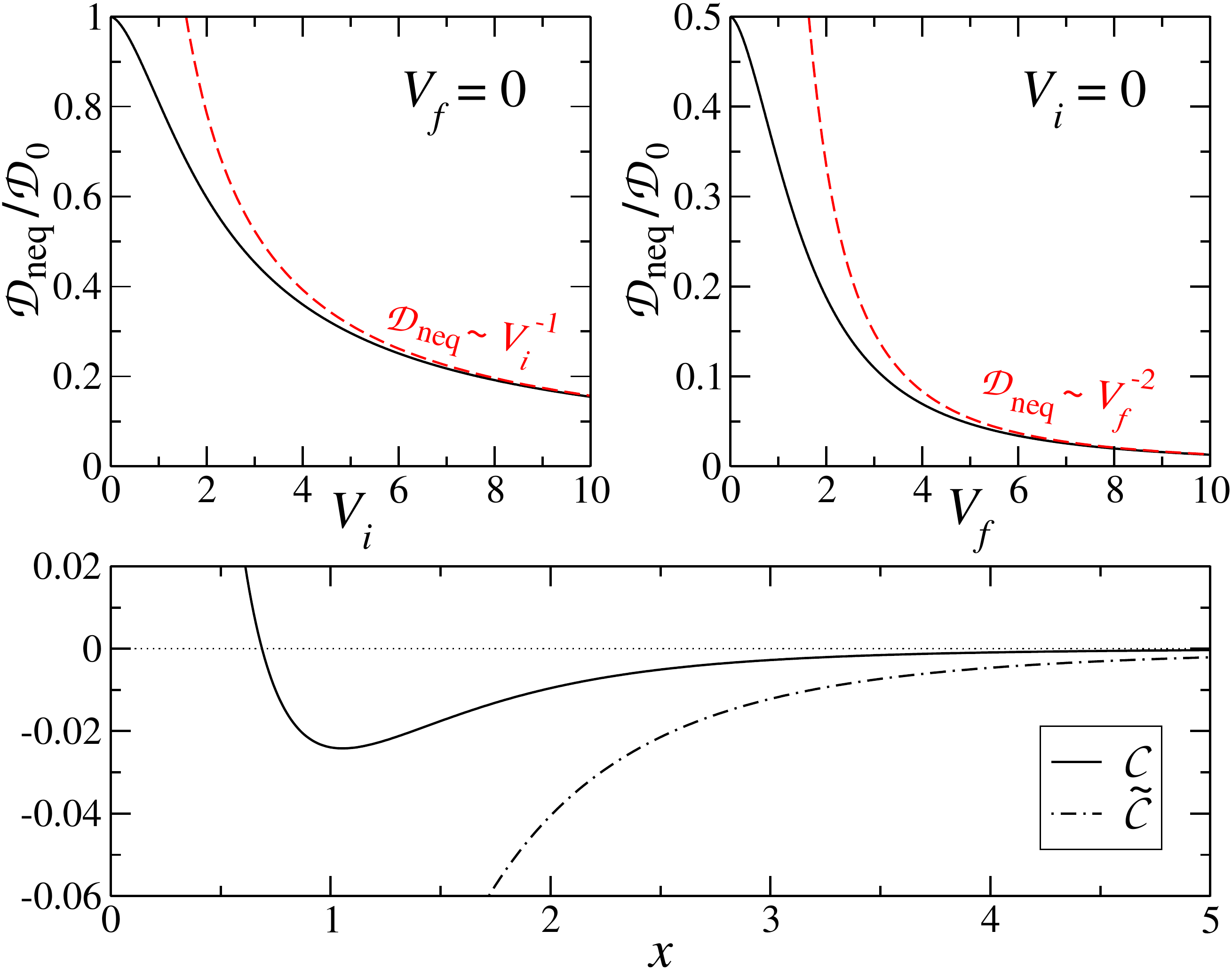}}
  \caption{(colour online). Average stiffness for model I, after a quench 
    $i)$ from the insulator at finite $V_i$, to $V_f=0$ (upper left panel);
    $ii)$ from the superfluid to finite $V_f$ (upper right panel).
    The stiffness is always smaller than that in the ground state at $V=0$,
    given by $\mathcal{D}_0 = 2t/\pi$.
    Red dashed lines denote the asymptotic behaviour at large $V$.
    Bottom panel: the functions $\mathcal{C}(x)$ (continuous line) 
    and $\tilde{\mathcal{C}}(x)$ (dashed line) appearing in the stiffness and in the
    helicity modulus for quenches within the insulator.}
\label{exactquench}
\end{figure}

We start considering the helicity modulus. For a quench from the superfluid into 
the insulator ${\cal Y}_{\rm neq}=0$, no matter the value of $V_f$.
Conversely, for a quench into the superfluid,
the helicity modulus coincides with the stiffness. 
It is then convenient to analyse directly ${\cal D}_{\rm neq}$. 
For a quench towards the superfluid (Fig.~\ref{exactquench}, upper left panel), 
as one could have guessed, the smaller is the quench,
the more the stiffness smoothly approaches the equilibrium value ${\cal D}_0$, 
while in the limit $V_i \gg 1$ it drops as ${\cal D}_{\rm neq} \sim t^2 / V_i$.
A similar behaviour occurs when quenching from 
the superfluid into the insulator (Fig.~\ref{exactquench}, upper right panel).
For $V_f \gg 1$ it scales as ${\cal D}_{\rm neq} \sim (4/3){\cal D}_0 (t/V_f)^2$;
the exact expression for any value of $V_f$ is given in~\cite{SI}.
A surprising situation arises for quenches within the insulator. 
Here, for small quenches, $\Delta V = \vert V_i - V_f \vert \ll 1$, the stiffness 
can be expressed as ${\cal D}_{\rm neq} = {\cal D}_0 \, \Delta V^2 \, \mathcal{C}(V_f)$,
while the helicity modulus contains an additional term:
${\cal Y}_{\rm neq} = {\cal D}_{\rm neq} + {\cal D}_0 \,\Delta V \,\tilde{\mathcal{C}}(V_f)$.
The functions $\mathcal{C}(x)$ and $\tilde{\mathcal{C}}(x)$ are plotted 
in Fig.~\ref{exactquench}, lower panel~\cite{SI}. 
Two things should be noticed. In contrast with the thermal case, 
the {\em stiffness} and the helicity modulus are {\em finite} and, in some cases, 
{\em negative}. This is a qualitative signature of the non-equilibrium 
steady state, and is not specific of this particular model.

Model II represents the simplest case of interacting bosons 
which cannot be mapped into free fermions. 
It is equivalent to a spin-$1/2$ XXZ Heisenberg chain.
We consider the antiferromagnetic region, $V \geq 0$, where a gapless superfluid ($V \leq 2$) 
is separated by a N\'eel state ($V > 2$). 
The stiffness has been investigated in finite-temperature equilibrium conditions, 
using analytical Bethe ansatz~\cite{Shastry_1990, Zotos_1999, Benz_2011, Prosen_2011, Prosen_2013}, 
as well as numerical techniques~\cite{Herbrych_2011, Moore_2012, HeidrichMeisner_2013}.
For the quenched case we address the whole spectrum of small systems (up to $L = 16$ sites) 
using direct numerical diagonalisation and then performing a finite-size scaling.

\begin{figure}[!t]
  \includegraphics[width=0.49\columnwidth]{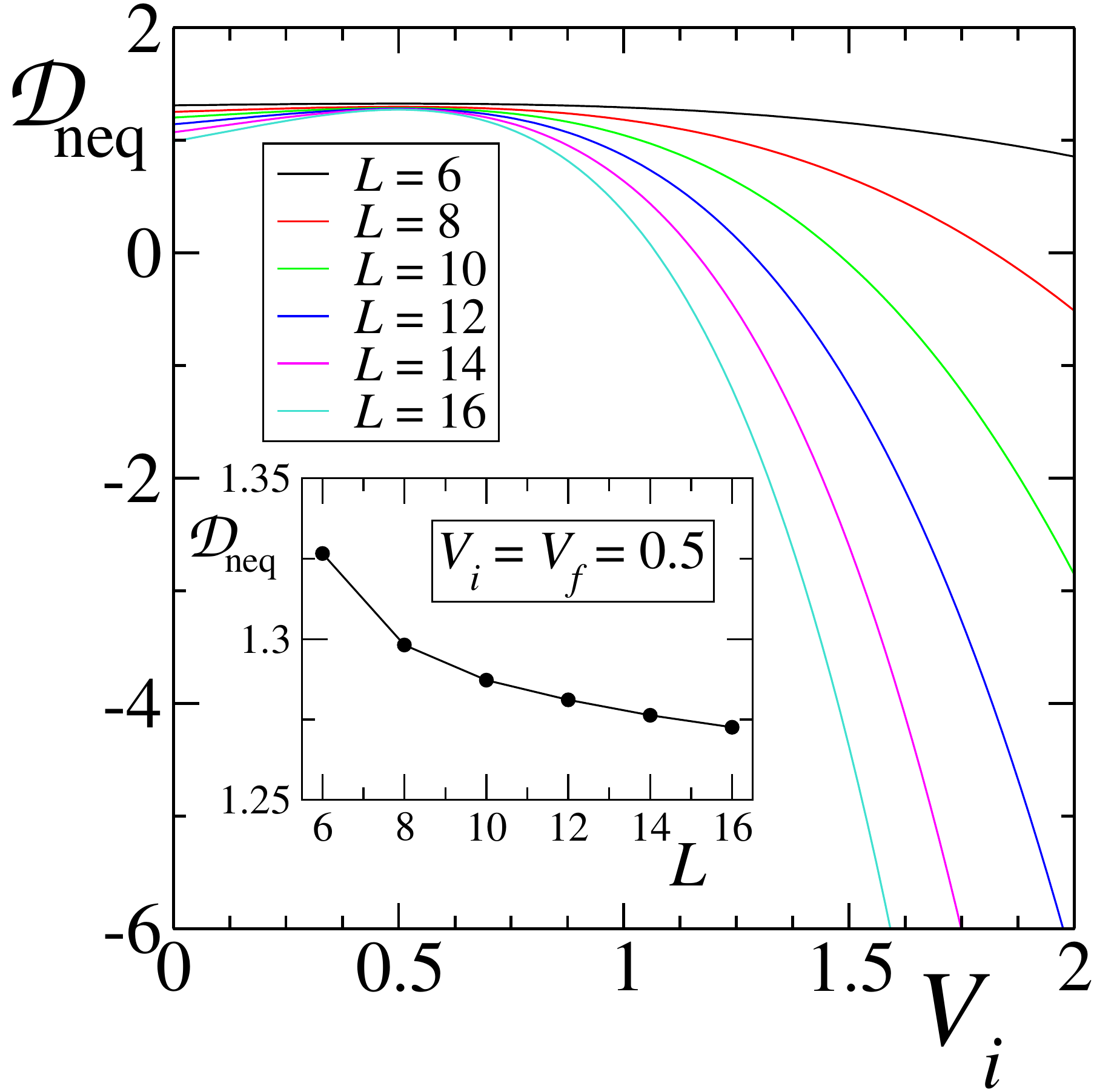}
  \includegraphics[width=0.49\columnwidth]{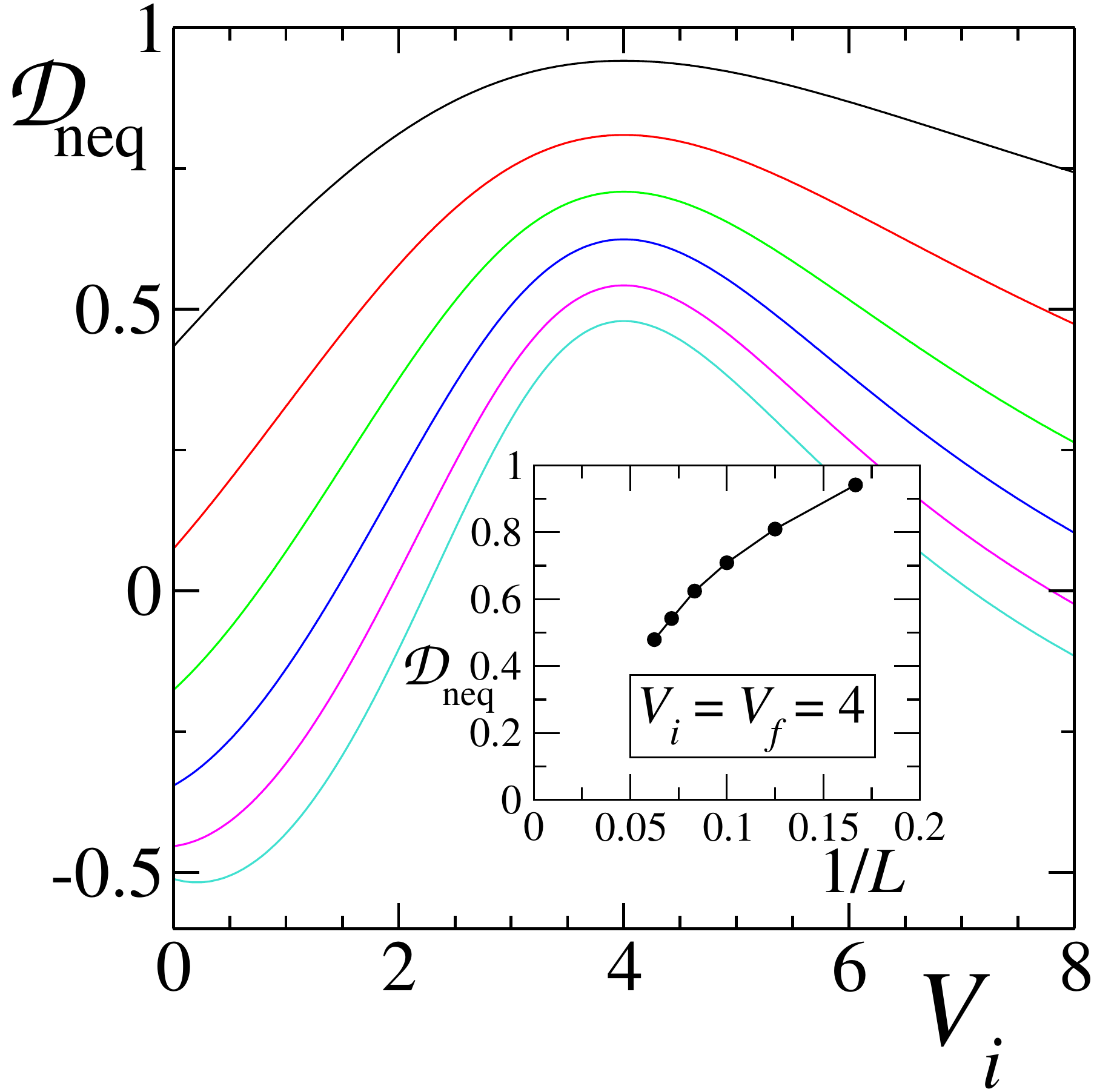}
  \caption{(colour online). Average stiffness for model II, 
    after a quench towards the superfluid ($V_f = 0.5$, left panel) 
    and towards the insulating phase ($V_f = 4$, right panel). 
    Different curves refer to various system sizes.
    Insets: stiffness in correspondence of $V_i = V_f$ (equilibrium cases).}
  \label{fig:xxz_INvar}
\end{figure}

We first discuss the stiffness for quenches from different initial states 
(different values of $V_i$) towards a fixed value of $V_f$. 
Fig.~\ref{fig:xxz_INvar} refers to quenches towards a superfluid (left panel) 
and an insulator (right panel). As in model I, 
in out-of-equilibrium conditions, ${\cal D}_{\rm neq}$ can assume negative values.
Moreover, we observe that the superfluid generally presents a stronger susceptibility
to phase twists, also in the excited levels. This appears from the sudden departure 
of ${\cal D}_{\rm neq}$ to large negative values in the left panel, when increasing $V_i$.
At equilibrium ($V_i = V_f$), after a finite-size scaling (insets),
one recovers for $L \to \infty$ a non-zero value only in the superfluid. 

\begin{figure}[!t]
  \includegraphics[width=0.9\columnwidth]{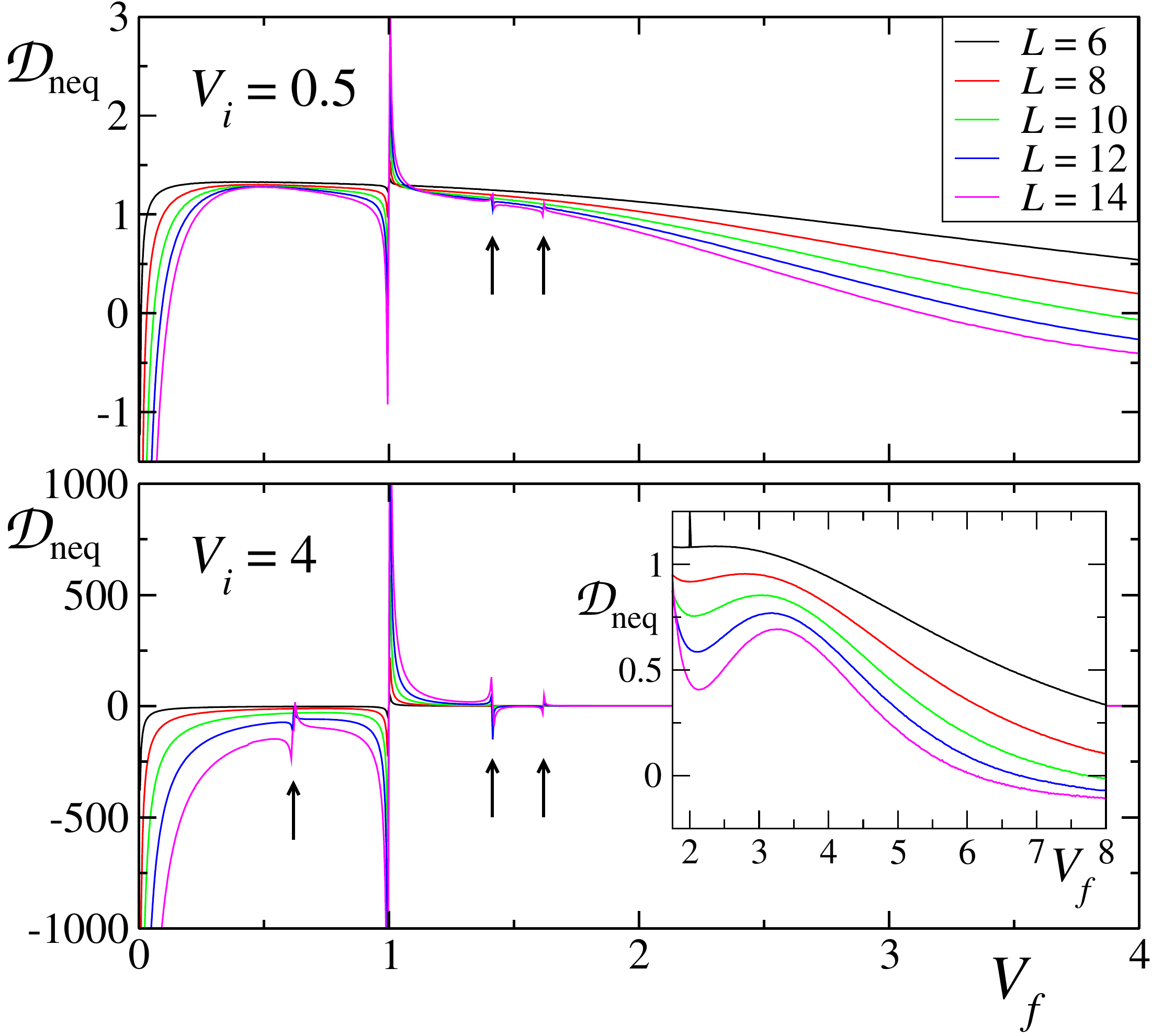}
  \caption{(colour online). Average stiffness for model II, 
    for a quench from the superfluid ($V_i = 0.5$, upper panel) 
    and from the insulating phase ($V_i = 4$, lower panels). 
    The arrows point out singular peaks emerging in correspondence 
    of $V^\star$ values obtained with $K \leq 5$ (together with $V^\star = 0,1,2$). 
    Inset: zoom of the region corresponding to quenches inside the insulator (with $V_i = 4$).} 
  \label{fig:xxz_FINvar}
\end{figure}

We conclude by analysing reverse quenches where 
the initial state is fixed and the final coupling varied. 
Results are shown in Fig.~\ref{fig:xxz_FINvar}. 
Different scales in the two panels reflect the enhanced sensitivity
to phase twists in the spectrum of a superfluid,
which better develops when the energy pumped by the quench is sufficiently
large to populate a consistent fraction of highly excited states
({\it i.e.}, when the initial state is insulating). 
For increasing system size, we observe the progressive appearance of a number 
of divergent peaks at special values of the interaction strength $V^\star$ in the superfluid.
In correspondence of these points an additional degeneracy given by 
the invariance under the loop algebra symmetry of $sl_2$ sets up,
causing a lack of completeness of the Bethe ansatz equations,
and is responsible for the occurrence of level crossings in response to a phase 
twist~\cite{Zotos_1999, Deguchi_2001, Prosen_2013}. 
They form a null set in the interval $0 \leq V_f \leq 2$, nonetheless become dense 
in the thermodynamic limit and are given by: $V^\star = q + q^{-1}$ with $q^{2K} = 1$, 
$K$ being a positive integer number~\cite{note_2}.
The divergences are caused by approaching the degeneracy points, 
where avoided crossings of levels are such that
$\lim_{V \to V^\star} \partial_\varphi^2 \varepsilon_n {\vert_{\varphi \to 0}} = \pm \infty$, 
and can lead to singularities in ${\cal D}_{\rm neq}$. 
Finally we stress that, for any $V_f \neq V^\star$, the whole excitation spectrum 
satisfies $\partial_\varphi \varepsilon_n {}_{\vert_{\varphi = 0}} = 0$, therefore
in model II ${\cal D}_{\rm neq} = {\cal Y}_{\rm neq}$, apart from 
the special points $V^\star$~\cite{note_1}.

The negative stiffness is a clear, qualitative and quantitative 
signature of the non-thermal nature of a steady state related 
to system integrability (or almost integrability). 
On the contrary, thermalisation in non-integrable systems should lead 
to a positive stiffness~\cite{note_3}. 
To corroborate this point, we considered numerically the effect of an integrability-breaking 
perturbation on model II (for a detailed analysis, see~\cite{SI}). 
When integrability is maximally broken, the stiffness from negative 
turns to be positive, confirming our expectations. 
Finally note that the negative stiffness does not result from the population 
of quasi-particles with negative temperature, in which case thermalisation 
to finite (positive) temperature states could never occur, independently of integrability. 
Rather in this case it follows from the fact that quasi-particles 
are populated non-thermally, yet not with a full population inversion. 

{\bf Experimental outlook. --- }
We showed that a study of the response 
to a twist in the boundary conditions provides distinct signatures 
of the non-equilibrium nature of the steady-state attained after a quench. 
Our results can acquire significance if they can be tested experimentally. 
All ingredients needed to measure ${\cal D}_{\rm neq}$ and ${\cal Y}_{\rm neq}$ 
are already present in ongoing experiments with cold bosons. 
Recent developments in manipulating toroidal Bose-Einstein condensates 
disclosed new tools to study persistent currents. 
They can be induced either by a rotating optical barrier~\cite{exp_nist} 
or by introducing artificial gauge fields~\cite{Cooper_2010, Dalibard_2011},
and already enabled the realization of an atomic superconducting 
quantum interference device~\cite{exp_lanl}. 
The necessary, albeit difficult, step is to perform a quantum quench in toroidal traps. 
Then a measure of the current flowing into the ring as a response to an external 
magnetic flux readily quantifies the stiffness (oscillating time-dependent flux) 
or the helicity modulus (constant flux). 
Finally we mention that bosons in optical lattices can simulate 
several magnetic models as, for example, the XXZ model studied here.

{\bf Acknowledgements. --- }
We acknowledge the EU (IP-SIQS), and the Italian
MIUR through PRIN (Project 2010LLKJBX) and FIRB (Project RBFR12NLNA) for financial support.

\newpage

\begin{widetext}

\appendix

\begin{center}
SUPPLEMENTARY MATERIAL
\end{center}

\section{Linear response after a quench}
\label{sec:LinResp}

The linear response theory to external perturbations can be easily generalised in the case of
quantum systems after a sudden quench.

Let us consider a given Hamiltonian $\hat{\Ham}_i$ which is suddenly quenched to $\hat{\Ham}_f$,
and is then subject to an external time-dependent perturbation $\hat{V}(t)$,
so that the total Hamiltonian after the quench is $\hat{\Ham}_f + \hat{V}(t)$.
The equation of motion for the state $\hat{\rho}(t)$ then reads:
\beq
   i \partial_t \hat{\rho}(t) - [ \hat{\Ham}_f, \hat{\rho}(t) ] = [ \hat{V}(t), \hat{\rho}(t)] \,.
   \label{eq:motion}
\eeq
Before continuing we observe that, without the perturbation $\hat{V}(t)$ 
and only in presence of the quench, one would have a state 
$\hat{\rho}_0(t) = e^{-i \hat{\Ham}_f t} \hat{\rho}_0(0) e^{i \hat{\Ham}_f t}$,
where the initial condition $\hat{\rho}_0(0) = \ket{\psi_0} \bra{\psi_0}$
is determined by the ground state $\ket{\psi_0}$ of $\hat{\Ham}_i$.
Such state would naturally evolve according to $\Hat{\Ham}_f$ and obey the equation 
of motion: $i \partial_t \hat{\rho}_0(t) = [ \hat{\Ham}_f, \hat{\rho}_0(t) ]$.
If we now add the perturbation $\hat{V}(t)$, we can split the various contributions
according to $\hat{\rho}(t) = \hat{\rho}_0 (t) + \delta \hat{\rho}(t)$.
Expliciting such expression in Eq.~\eqref{eq:motion}, we get
\beq
   \left\{ i \partial_t \hat{\rho}_0(t) - [ \hat{\Ham}_f, \hat{\rho}_0(t) ] \right\} +
   \left\{ i \partial_t \delta \hat{\rho}(t) - [ \hat{\Ham}_f, \delta \hat{\rho}(t) ] \right\}
   = [ \hat{V}(t), \hat{\rho}(t)] \,.
   \label{eq:motion2}
\eeq
Now, as stated above, the two terms in the first curl brackets constitute the  
equation of motion for $\hat{\rho}_0(t)$, therefore their sum is equal to zero,
while the rest dictates the evolution for the remaining part $\delta \hat{\rho}(t)$.
We then observe that, by means of simple algebraic passages, 
the terms in the second curl brackets can be rewritten according to
$e^{-i \hat{\Ham}_f t} \big( i \partial_t [e^{i \hat{\Ham}_f t} \, \delta \hat{\rho}(t) \, 
e^{-i \hat{\Ham}_f t}] \big) e^{i \hat{\Ham}_f t}$, therefore Eq.~\eqref{eq:motion2} 
leads us to a simple equation of motion for $\delta \hat{\rho}(t)$:
\beq
   i \partial_t \delta \hat{\tilde{\rho}}(t) = 
   e^{i \hat{\Ham}_f t} \, [ \hat{V}(t), \hat{\rho}(t)] \, e^{-i \hat{\Ham}_f t} \,,
   \label{eq:Kubo_int}
\eeq
where we used the interaction picture, according to which
$\delta \hat{\tilde{\rho}} = e^{i \hat{\Ham}_f t} \delta \hat{\rho} \, e^{-i \hat{\Ham}_f t}$.
If we finally integrate this equation and then insert back the definition 
of $\hat{\tilde{\rho}}(t)$, we get
\beq
   \delta \hat{\rho}(t) \simeq - i e^{-i \hat{\Ham}_f t} 
   \left\{ \int_0^t [ \hat{\tilde{V}}(t'), \hat{\rho}(0)] \, {\rm d} t' \right\} e^{i \hat{\Ham}_f t} \,,
   \label{eq:Rhot_kubo0}
\eeq
where we used again the notation 
$\hat{\tilde{A}}(t) = e^{i \hat{\Ham}_f t} \, \hat{A} \, e^{-i \hat{\Ham}_f t}$, for a generic operator $\hat{A}$.
To get the previous equation we also approximated $\hat{\rho}(t)$ 
in the r.h.s. of Eq.~\eqref{eq:Kubo_int} with $\hat{\rho}_0(t)$,
and used $\delta \hat{\tilde{\rho}}(0) = 0$, since $\hat{\rho}(0) = \hat{\rho}_0(0)$.

Equation~\eqref{eq:Rhot_kubo0}, together with the fact that the full density matrix
of the system is given by $\hat{\rho}(t) = \hat{\rho}_0 (t) + \delta \hat{\rho} (t)$, 
leads us to the generalisation of the Kubo formula for the quenched case.
The expectation value of any given operator 
$\langle \hat{A} \rangle (t) = {\rm Tr} [\hat{\rho}(t) \hat{A}]$ is given by: 
\beq
   \langle \hat{A} \rangle (t) =  \langle \hat{\tilde{A}}(t) \rangle_{0} 
   - i \int_0^t \big\langle [ \hat{\tilde{A}}(t), \hat{\tilde{V}}(t') ] \big\rangle_{0} \, {\rm d}t' \,,
      \label{eq:Rhot_kubo}
\eeq
where $\langle \, \cdot \, \rangle_{0} = {\rm Tr} [\hat{\rho}(0) \, \cdot\, ]$ denotes 
the unperturbed expectation value, {\it i.e.}, the equilibrium average with respect to $\hat{\Ham}_i$.

Let us now take a generic time-dependent perturbation $\hat{V}(t) = h(t) \, \hat{B}$.
From Eq.~\eqref{eq:Rhot_kubo} we obtain thus the generalised Kubo formula for a quenched system:
\beq
   \langle \hat{A} \rangle (t) =  \langle \hat{A}(t) \rangle_{0} 
   + \int_0^t {\rm d}t' \, \chi_{AB}(t,t') \, h(t') \,, \quad {\rm with} \qquad
   \chi_{AB}(t,t') = -i \, \theta(t-t') \, \big\langle [ \hat{A}(t), \hat{B}(t') ] \big\rangle_{0} \,,
\eeq
where the response function $\chi_{AB}(t,t')$ now depends explicitly from $t$ and $t'$ separately.
In the previous equation we also omitted the tildes on the operators, since from now on, 
time-dependent operators will be understood in the Heisenberg representation.

In an out-of-equilibrium process, to get expectation values of time-dependent quantities 
it is natural to perform a time average for long times.
For this reason, in this context it is useful to introduce the Wigner coordinates 
$T = (t + t')/2$ and $\tau = t - t'$, and to perform averages according to 
$\overline{[ \, \cdot \,]} = \lim_{T \to \infty} \frac{1}{T} \int_0^T [ \, \cdot \,]$.
In this way one obtains an expression for the averaged response function in terms
of difference of times $\tau$:
\beq
   \overline{\chi_{AB}(T,\tau)} = -i \, \theta(\tau) \, \lim_{T \to \infty} \frac{1}{T}
   \int_0^T {\rm d}t' \; {\rm Tr} \bigg\{ \hat{\rho}(0) \cdot \Big[ \hat{A} \Big( T + \frac{\tau}{2} \Big), 
     \hat{B} \Big( T - \frac{\tau}{2} \Big) \Big] \bigg\} \,,
   \label{eq:timeAvg}
\eeq

Without loss of generality, let us now specialise to the case of pure initial states 
$\hat{\rho}(0) = \ket{\psi_0} \bra{\psi_0}$,
where $\ket{\psi_0}$ is the ground state of the initial Hamiltonian $\hat{\Ham}_i$.
The previous expression in Eq.~\eqref{eq:timeAvg} can then be further simplified 
by noticing that:
i) the commutator inside brackets can be rewritten as
$e^{-i \hat{\Ham}_f T} \big[ \hat{A}(\tau/2), \hat{B}(-\tau/2) \big] e^{i \hat{\Ham}_f T}$;
ii) under the average over $T$, time-oscillating terms can be neglected,
and the trace over $\hat{\rho}(0)$ can be performed on the diagonal ensemble 
$\hat{\rho}_{\rm st} = \overline{\hat{\rho}(t)} = \sum_{n,r} p_n \ket{\psi_{n,r}} \bra{\psi_{n,r}}$.
Here the weights of the diagonal ensemble, $p_n = \bra{\psi_0} \hat{\pi}_n \ket{\psi_0}$, 
are the probabilities to find the initial state in the $n$-th eigenspace
of $\hat{\Ham}_f$ (this is characterised by the eigenstates $\ket{\psi_{n,r}}$,
$r$ being a degeneracy index).
We eventually get
\beq
   \overline{\chi_{AB}(\tau)} = -i \, \theta(\tau) \, 
   \sum_{n,r}  p_n \bra{\psi_{n,r}} \big[ \hat{A}(\tau/2), \hat{B}(-\tau/2) \big] \ket{\psi_{n,r}} \,.
\eeq
which coincides with Eq.~(1) in the main text.

\section{Operational protocols to probe superfluidity after a quench}
\label{sec:Quench+LinResp}

In order to address the non-equilibrium physics of a many-body quantum system, 
we adopt the simplest conceivable setting where to study its relaxation: 
a sudden quench of one of the control parameters $\lambda$ of the corresponding Hamiltonian.
Namely, we suppose to work at zero temperature and to perform, at a certain 
reference time $t_0 = 0$, the instantaneous change $\lambda_i \to \lambda_f$.
Such kind protocol is routinely performed in out-of-equilibrium experiments 
with cold atoms, even with superimposed optical lattices faithfully mimicking 
the physics described by the type of models in Eq.~(5) of the main text.
This is schematically shown in Fig.~\ref{fig:Quench+LinResp}, black line.

\begin{figure}[!b]
  \includegraphics[width=0.4\columnwidth]{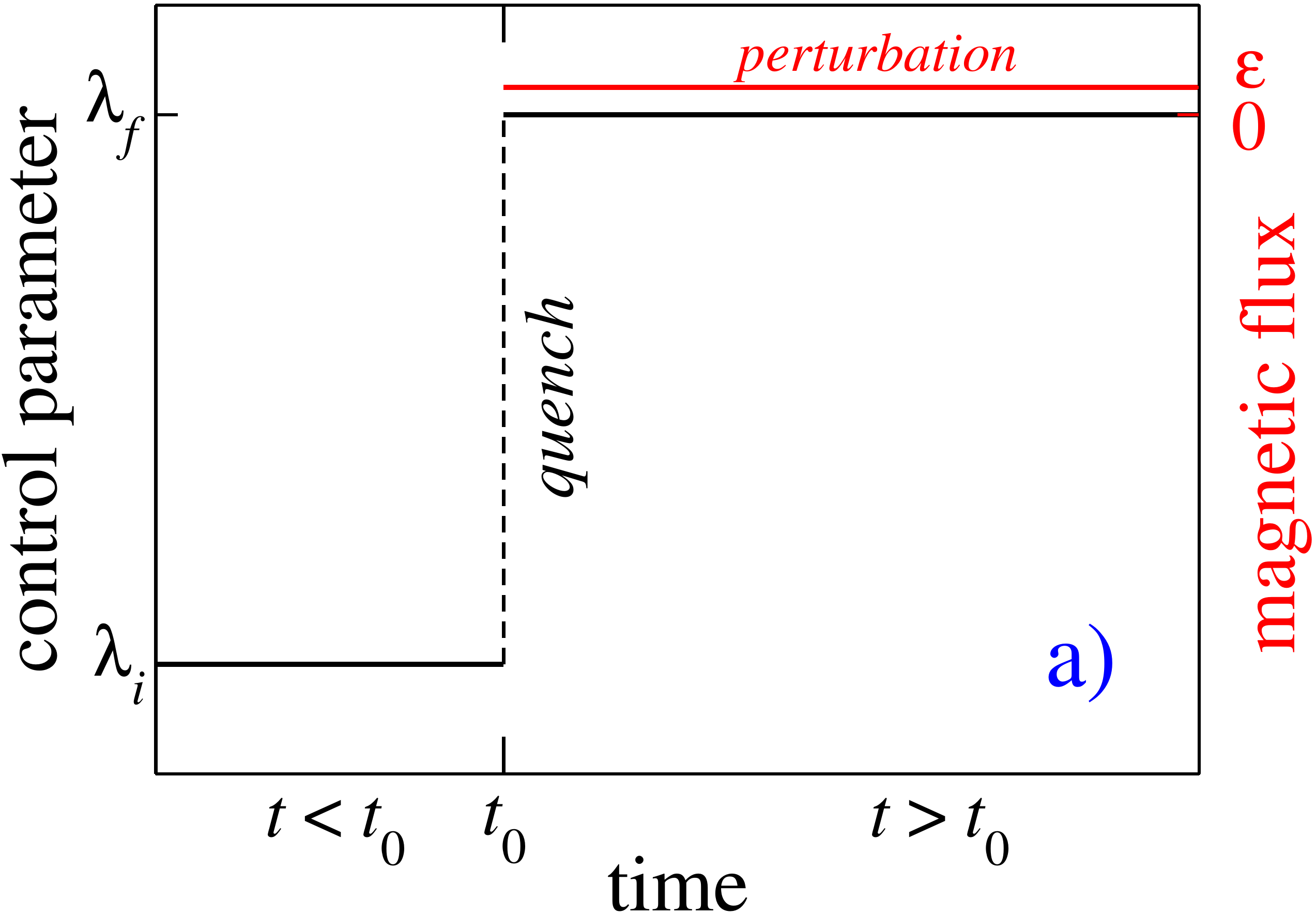} \hspace{1.5cm}
  \includegraphics[width=0.4\columnwidth]{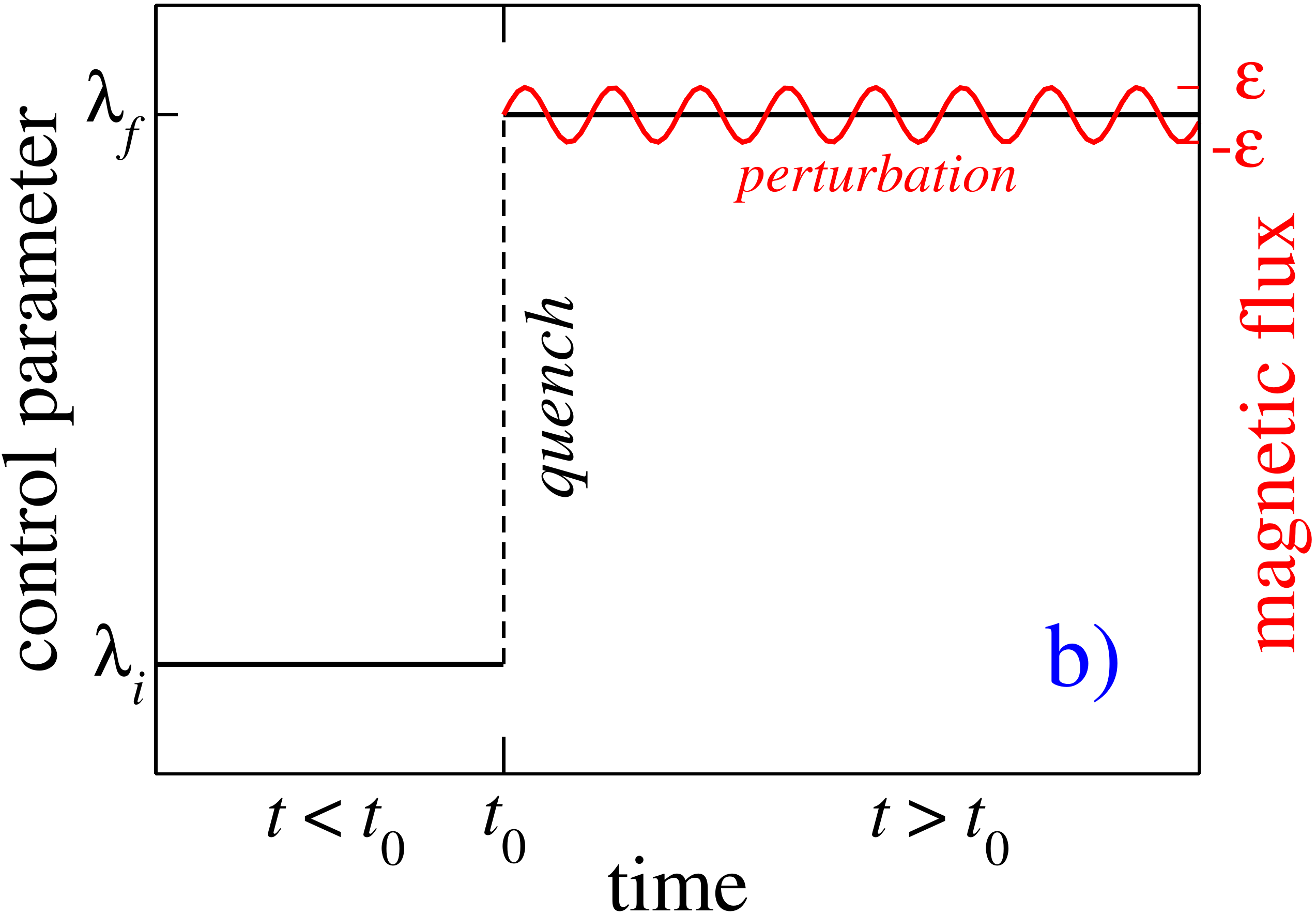}
  \caption{(colour online). Prototypical experiments to study the system's 
    superfluidity following a quantum quench of the parameter $\lambda$ 
    at time $t_0$, according to linear response. 
    The perturbation, {\it i.e.}, an external and small magnetic flux piercing the ring
    here depicted in red, has to be switched on after the quench (for $t>t_0$).
    Panel {\bf a}: The helicity modulus quantifies the susceptibility
    to a weak and static magnetic flux.
    Panel {\bf b}: The charge stiffness characterises the
    response to a weak and time-dependent oscillating flux.}
  \label{fig:Quench+LinResp}
\end{figure}

The paradigmatic way to measure superfluidity at equilibrium goes through
the use of periodic boundary conditions (a ring in the one-dimensional scenario).
An analogous geometry has to be implemented in suitable out-of-equilibrium experiments.
Despite the apparent difficulty in combining and controlling these two ingredients
(non-equilibrium and ring-shaped systems), the recent advances in the manipulation 
of cold atoms with toroidal magneto-optical traps have disclosed 
novel valid strategies to achieve this goal in the near future.

Once the quench in a toroidal trap has been realised, the superfluidity
of the stationary state can be probed by adopting a protocol which is formally
equivalent to the standard one performed at equilibrium.
The fundamental quantity to be measured is the persistent current that is establishing 
into the ring in response to a magnetic flux piercing the system.
This is equivalent to an electric field parallel to the chain, which 
can be conveniently described, after performing a Peierls substitution, 
by a complex phase acquired by the hopping amplitude.

To be precise, one needs to switch on a magnetic field passing through the toroidal
trap, for times $t > t_0$. Its shape is depicted in Fig.~\ref{fig:Quench+LinResp}, red line.
As a function of time, this field can be static or oscillating with
a certain frequency $\omega$.
These express two operatively different protocols to measure superfluidity,
according to the two quantities detailed in the main text.

$\bullet$ The helicity modulus is the derivative of the time-averaged persistent current 
that asymptotically establishes in the ring at long times, with respect to an infinitesimal 
and constant magnetic flux density (panel {\bf a}).

$\bullet$ The stiffness is the zero-frequency Drude weight in the conductivity $\sigma$
with respect to a sinusoidal monochromatic electric field. 
This is ${\cal D}_{\rm neq} = \frac{\omega}{2} \, {\rm Im} [\sigma(\omega) ] \vert_{\omega \to 0}$ 
(panel {\bf b}).

We stress persistent currents for Bose Einstein condensates in toroidal traps have been 
already measured, while establishing this kind of experiments in a quench scenario 
constitutes a novel challenge which could be addressed soon
with already available techniques.

\section{Hard-core bosons in a staggered field}
\label{sec:Model1}

In this section we focus on the Hamiltonian describing Model I in the main text, {\it i.e.},
\beq
   \hat{\Ham}(\varphi) = 
   t \sum_{j=1}^L \left( e^{i \varphi} \hat{b}^\dagger_{j} \hat{b}_{j+1} 
   + e^{-i\varphi} \hat{b}^\dagger_{j+1} \hat{b}_j \right)
   + V \sum_{j=1}^L (-1)^j \hat{b}^\dagger_j \hat{b}_j \,.
   \label{eq:model}
\eeq
Here $\hat{b}^\dagger_j / \hat{b}_j$ denote hard-core bosonic creation/annihilation operators 
on site $j$, while $-t$ and $V$ respectively denote the nearest-neighbour hopping strength 
and the intensity of an external staggered magnetic field. 
The system has a finite length $L$ and is supposed to have periodic boundary 
conditions ($\hat{b}_{L+1} = \hat{b}_1$). $\varphi$ is the flux passing through such ring.
In the following we will analyse the behaviour of the superfluid stiffness and of
the helicity modulus for $L \to \infty$, thus recovering the asymptotic behaviour
at the thermodynamic limit.

The Hamiltonian~\eqref{eq:model} can be readily mapped into a free-fermion model 
by first employing a Jordan-Wigner (JW) transformation of hard-core bosons into fermions:
\beq
   \hat{b}^\dagger_j = \exp \Big( i \pi \sum_{l<j} \hat{a}^\dagger_l \hat{a}_l \Big) \hat{a}^\dagger_j \, ,
\eeq
where $\hat{a}^\dagger_j / \hat{a}_j$ are fermionic creation/annihilation operators, satisfying
canonical anticommutation relations. 
In this way, one gets straightforwardly an expression for the new Hamiltonian that is 
formally identical to Eq.~\eqref{eq:model}, but with the new JW fermions.
After switching to momentum representation, by Fourier transforming the fermionic operators
\beq
   \hat{a}_j = \frac{1}{\sqrt{L}} \sum_k e^{-ikj} \hat{a}_k \,,
\eeq
where $k = \pm \pi (2n+1)/L$ and $n=0, \ldots, L/2-1$,
the Hamiltonian can be rewritten according to the following compact form: 
\beq
   \hat{\Ham}(\varphi) = 
   \sum_{\vert k \vert < \pi/2} \hat{\vec{\Psi}}^\dagger_k \, \hat{H}_k(\varphi) \, \hat{\vec{\Psi}}_k \,,
   \label{eq:quadratic}
\eeq
with $\hat{\vec{\Psi}}_k = \big( \hat{a}_k \;,\; \hat{a}_{k+\pi} \big)^T$ and
$\hat{H}_k(\varphi) = 2 t \cos(k-\varphi) \hat{\sigma}^z + V \hat{\sigma}^x$. 
Here $\hat{\sigma}^\alpha$ ($\alpha=x,y,z$) are the spin-1/2 Pauli matrices,
while momenta $k$ are defined $[{\rm mod} (2 \pi) ]$.
The quadratic form~\eqref{eq:quadratic} is then diagonalised by means of a Bogoliubov rotation: 
\beq
   \hat{\vec{\Gamma}}_k = e^{i \, \Theta_k(\varphi) \hat{\sigma}^y} \hat{\vec{\Psi}}_k
   \label{eq:Bogo1}
\eeq
of an angle $\Theta_k(\varphi)$ defined by
\beq
   \tan \big[ 2 \, \Theta_k(\varphi) \big] = \frac{V}{2t \cos(k-\varphi)} \,,
   \label{eq:Bogo2}
\eeq
which maps $\hat{\vec{\Psi}}_k$ into 
$\hat{\vec{\Gamma}}_k = \big( \hat{\gamma}_k^+ \;,\; \hat{\gamma}_k^- \big)^T$,
so that
\beq
   \hat{\Ham}(\varphi) = \sum_{\vert k \vert < \pi/2} 
   \epsilon_k(\varphi) \; \hat{\vec{\Gamma}}^\dagger_k \, \hat{\sigma}^z \, \hat{\vec{\Gamma}}_k 
   \qquad {\rm with} \quad 
   \epsilon_k(\varphi) = \sqrt{ \big[ 2 t \cos (k - \varphi) \big]^2 + V^2} \,.
   \label{eq:HamDiag}
\eeq
This implies that the system eigenmodes can be grouped in couples of momentum $k$,
which have the same absolute energy $\epsilon_k$ and opposite signs,
{\it i.e.}, with energy $\varepsilon^\pm_{k}(\varphi) = \pm \epsilon_k(\varphi)$.

Equation~\eqref{eq:HamDiag} allows us to compute both the stiffness $\mathcal{D}$
and the helicity modulus $\mathcal{Y}$ in terms of single-particle quantities.
We start from the time-averaged current density, which is defined as:
\beq
   \overline{j_\varphi} = - \frac{1}{L} \lim_{T \to \infty} \frac{1}{T} 
   \int_0^T {\rm d}t \bra{\psi(t)} \partial_\varphi \hat{\Ham} \ket{\psi(t)} \,,
\eeq
where $\ket{\psi(t)}$ is the system wavefunction at time $t$.
In the long-time limit one can drop the rotating terms and keep only the ``diagonal''
contributions, thus obtaining: 
\beq
   \overline{j_\varphi} = -\frac{1}{L} \sum_{\vert k \vert < \pi/2} \, 
   \sum_{i=\pm} p^i_k(\varphi) \, \partial_\varphi \varepsilon^i_k(\varphi) \, ,
\eeq
with $p_k^i = \bra{\psi_0} \hat{\gamma}_k^i {}^\dagger \, \hat{\gamma}_k^i \ket{\psi_0}$ 
being the overlap between the state $\ket{\psi}$ at any time and the $k$-th eigenmode 
$\hat{\gamma}_k$.
The helicity modulus $\mathcal{Y} = - \partial_\varphi \overline{j_\varphi} {}_{\vert_{\varphi \to 0}}$
is thus given by:
\beq
   \mathcal{Y} = \mathcal{D} + \frac{1}{L} \sum_{\vert k \vert < \pi/2} \,
   \sum_{i = \pm} \big[ \partial_\varphi p_k^i (\varphi) \big] \,
   \big[ \partial_\varphi \varepsilon_k^i(\varphi) \big]_{\vert_{\varphi \to 0}}
   \label{eq:helic}
\eeq
where $\mathcal{D}$ denotes the stiffness:
\beq
   \mathcal{D} = 
   \frac{1}{L} \sum_{\vert k \vert < \pi/2} \, \sum_{i=\pm} p_k^i(\varphi) \,
   \partial^2_\varphi \varepsilon_k^i(\varphi) {}_{\vert_{\varphi \to 0}} \,.
   \label{eq:stiff}
\eeq
This expression for the stiffness can be rewritten according to 
$\mathcal{D} = \mathcal{D}_0 + \delta \mathcal{D}$, with
\beq
   \mathcal{D}_0 = 
   - \frac{1}{L} \sum_{\vert k \vert < \pi/2} \partial^2_\varphi \epsilon_k(\varphi) {}_{\vert_{\varphi \to0}} \; ,
   \qquad
   \delta \mathcal{D} = \frac{2}{L} \sum_{\vert k \vert < \pi/2} 
   p_k^+ (\varphi) \, \partial^2_\varphi \epsilon_k(\varphi){}_{\vert_{\varphi \to 0}}  \,,
   \label{eq:stiff_fin}
\eeq
where $\mathcal{D}_0$ denotes the ground-state contribution to the stiffness,
while $\delta \mathcal{D}$ contains all the corrections due to finite temperature, or to a quench.
Here we used the fact that $p_k^+(\varphi) + p_k^-(\varphi) = 1$.

\subsection{Equilibrium} \label{sec:equil}

\subsubsection*{Zero-temperature ground state ($T=0$)}

In this case $\ket{\psi}$ coincides with the ground state $\ket{\psi_0}$, 
which is trivially written in the Fock space of the eigenmodes $\hat{\gamma}^\pm_k$.
Starting from the vacuum state for the JW fermions $\ket{0}$
and neglecting normalisation constants, this is simply given by:
$\ket{\psi_0} = \prod_{\vert k \vert < \pi/2} \hat{\gamma}^-_k {}^\dagger \ket{0}$,
and its energy is:
\beq
   E_0(\varphi) = \sum_{\vert k \vert < \pi/2} \varepsilon_k^- (\varphi)
   = -\frac{L}{2\pi} \int_{-\pi/2}^{\pi/2} {\rm d}k \, \sqrt{ \big[2t \cos(k-\varphi) \big]^2 + V^2} \,.
   \label{eq:E0_A}
\eeq
Note that here the discrete sum over the momenta $k$ has been replaced by an integral, 
in order to work in the thermodynamic limit $L \to \infty$
(hereafter we will always perform calculations in this limit). 
This expression can be recast in terms of the complete elliptic integral 
of the second kind $\mathbb{E}(x) \equiv \mathbb{E}(\frac{\pi}{2} \vert x)$, with
\beq
   \mathbb{E}(\phi \vert m) = \int_0^\phi (1 - m \sin^2 \theta)^{1/2} {\rm d}\theta \,,
   \label{eq:elliptic2}
\eeq
according to the following:
\beq
   E_0(\varphi) = -\frac{L}{\pi} \sqrt{4 t^2 + V^2} \;\, 
   \mathbb{E} \bigg( \! \frac{4 t^2}{4 t^2 + V^2} \! \bigg)\,.
   \label{eq:E0_eq}
\eeq
To get this expression, we used the property $\mathbb{E}(\frac{\pi}{2}+\varphi \, \vert \, x) 
+ \mathbb{E}(\frac{\pi}{2}-\varphi \, \vert \, x) = 2 \, \mathbb{E}(x)$.
Equation~\eqref{eq:E0_eq} readily implies that $\mathcal{D} = \mathcal{Y} = 0$ for any $V>0$,
consistently with the fact that the model is a gapped insulator there.
The case $V=0$ requires a separate treatment, since in this case Eq.~\eqref{eq:E0_A} becomes
\beq
   E_0(\varphi) = -\frac{2tL}{\pi} \cos \varphi \,,
\eeq
from which it follows that $\mathcal{D} = \mathcal{Y} = 2t/ \pi$ at the superfluid point $V=0$.

\subsubsection*{Finite-temperature thermal state ($T>0$)}

Since the diagonalised Hamiltonian~\eqref{eq:HamDiag} is written in terms of free fermions,
the occupation probabilities $p_k^i(\varphi)$ at equilibrium and at finite inverse
temperature $\beta = 1/T$ (hereafter we are taking $k_B = 1$)
are given by the standard Fermi distribution function:
\beq
   p_k^\pm(\varphi) = \big[ e^{\pm \beta \epsilon_k(\varphi)} + 1 \big]^{-1} \,.
   \label{eq:Fermi}
\eeq

$\bullet$ Let us start with the evaluation of the stiffness, by inserting the Fermi function
in Eq.~\eqref{eq:stiff}, so that we get:
\beq
   \mathcal{D} = - \frac{1}{L} \sum_{\vert k \vert < \pi/2} \tanh \bigg( \frac{\beta \epsilon_k}{2} \bigg) \,
   \partial^2_\varphi \epsilon_k(\varphi) {}_{\vert_{\varphi \to 0}} \, .
   \label{eq:stiff_T>0}
\eeq

Also in this case, it is useful to distinguish among the insulating and the superfluid.
At $V=0$, one has $\epsilon_k(\varphi) = 2t \cos(k-\varphi)$ and therefore
$\partial^2_\varphi \epsilon_k(\varphi) {}_{\vert_{\varphi \to 0}} = -2 t \cos(k)$.
Inserting it into Eq.~\eqref{eq:stiff_T>0},
we get an integral expression for the stiffness which can be numerically worked out.
Comparing the temperature with the bandwidth, in the limit $T \gg t$ one can
use the expansion $\tanh (x) \stackrel{x \to 0}{\approx} x$ 
and thus get $\mathcal{D} \approx t^2 / 2 T$.
In the more relevant limit $T \ll t$ of small temperature, one finds 
$\mathcal{D} \approx \mathcal{D}_0 \big[ 1 - (\pi^2/24) (T/t)^2 \big]$,
where we defined $\mathcal{D}_0 = 2 t / \pi$ as the maximum reachable value 
by the stiffness at zero temperature.
We now concentrate on the case $V \neq 0$ and focus on temperatures below the bandwidth.
Assuming $T \ll t$ and also $t > V$ one can show that
\beq
   \delta \mathcal{D} \approx \sqrt{\frac{8}{\pi}} \, t \, \sqrt{\frac{T}{V}} \, e^{-V/T} \,,
\eeq
so that the stiffness approaches the zero-temperature value $\mathcal{D}_0 = 0$ exponentially.

Two plots for the finite-temperature stiffness in the thermal state,
in the cases $V=0$ and $V \neq 0$, are shown in Fig.~\ref{fig:stiff_temp}.
The analytic estimates are plotted in colored dashed lines and are in good agreement
with the numerics in the respective low-temperature (blue) and high-temperature (red) regime.

\begin{figure}
  \includegraphics[width=0.7\columnwidth]{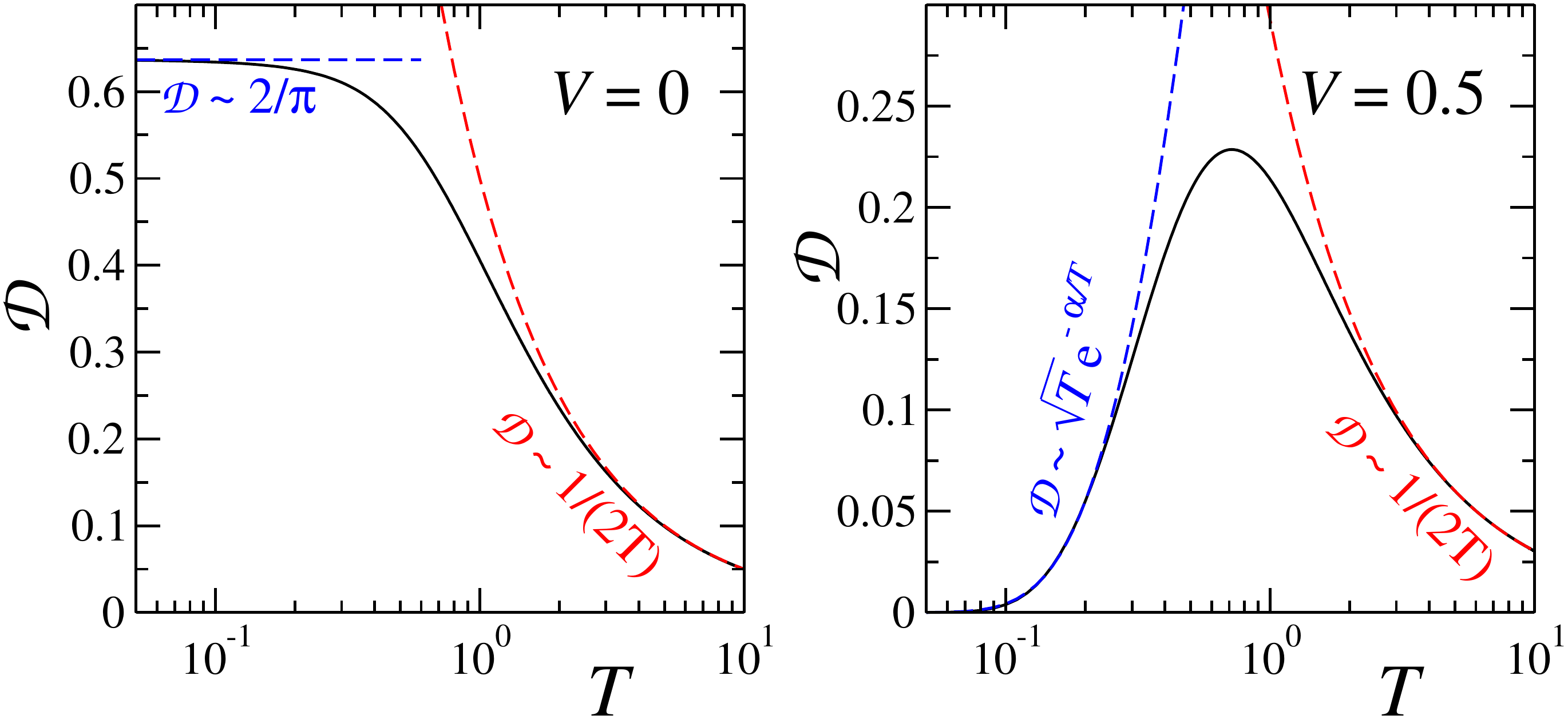}
  \caption{(colour online) Stiffness as a function of the temperature, 
    for $V=0$ (left panel, superfluid at $T=0$)
    and for $V=0.5$ (right panel, insulator at $T=0$).
    Dashed lines indicate the analytic estimates available for small (dashed blue lines)
    and for large (dashed red lines) temperature (see the text).
    Here we have set $t=1$ as the energy scale of the system.}
  \label{fig:stiff_temp}
\end{figure}

$\bullet$ Finally we consider the helicity modulus
$\mathcal{Y} = \frac{1}{L} \sum_k \sum_i \partial_\varphi 
\big[ p_k^i (\varphi) \, \partial_\varphi \varepsilon_k^i(\varphi) \big] {}_{\vert_{\varphi \to 0}}$.
Since the expression in square brackets depends on the flux $\varphi$ 
always in the combination $k - \varphi$ [see Eqs.~\eqref{eq:HamDiag} and~\eqref{eq:Fermi}], 
when taking the outer derivative one can first change variable $\varphi \to k$,
and then carry out the summation (integration in the continuum limit) over $k$ 
by evaluating such expression at its extrema:
\beq
   \mathcal{Y}=\frac{1}{2\pi} \sum_{i=\pm} p_k^i \, \partial_k \varepsilon_k^i \Big\vert_{-\pi/2}^{\pi/2} \,.
   \label{eq:helic_fin}
\eeq
For any $V \neq 0$ we have that $\partial_k \varepsilon_k^\pm {}_{\vert_{k = \pi/2}} = 0$, 
therefore $\mathcal{Y} = 0$.
On the other hand, for $V=0$ we have $\partial_k \varepsilon_k {}_{\vert_{k = \pm \pi/2}} = \pm 2t$ 
and $p^\pm_{\pm \pi/2} = 1/2$, hence $\mathcal{Y} = \mathcal{D}_0$.
This shows that, at finite temperature, stiffness and helicity modulus {\it do not coincide}
for this model.

\subsection{Out of equilibrium}

All the previous calculations for the helicity modulus and the stiffness can be 
easily generalised to the case of a sudden quench.
In particular, at a given time $t_0$ we suppose to perform a quench 
of the parameter $V$: $V_i \to V_f$ and simultaneously add a magnetic flux 
$\varphi$ through the ring: $0 \to \varphi$.

Before proceeding, we have to compute the probabilities $p_k^\pm(\varphi)$.
The two Hamiltonians corresponding to the situation before and after the quench
can be both diagonalised through a Bogoliubov transformation of the type 
in Eqs.~\eqref{eq:Bogo1} and~\eqref{eq:Bogo2}, {\it i.e.}, 
$\hat{\vec{\Gamma}}_k(V,\varphi) = e^{i \, \Theta_k(V, \varphi) \hat{\sigma}^y} \hat{\vec{\Psi}}_k$.
For the sake of clarity, from now on, when necessary we will explicitly indicate 
the dependence of the various quantities from $V$ and $\varphi$.
Therefore, for a generic quench $(V_i, 0) \to (V_f, \varphi)$ we have:
\beq
   \hat{\vec{\Gamma}}_k (V_f, \varphi) =
   \big[ \cos (\Delta \Theta_k) +i \sigma^y \sin (\Delta \Theta_k) \big] \, \hat {\vec{\Gamma}}_k(V_i,0)
   \qquad {\rm with} \quad \Delta \Theta_k \equiv \Theta_k(V_f, \varphi) - \Theta_k(V_i, 0) \,.
   \label{eq:GammaQuench}
\eeq
By definition, the ground state of the initial Hamiltonian is such that
$\langle \hat{\gamma}^+_{k, \,(V_i,0)} {}^\dagger  \, \hat{\gamma}^+_{k, \,(V_i,0)} \rangle = 
\langle \hat{\gamma}^-_{k, \,(V_i,0)} {}^\dagger  \, \hat{\gamma}^+_{k, \,(V_i,0)} \rangle = 0$
and $\langle \hat{\gamma}^-_{k, \,(V_i,0)} {}^\dagger  \, \hat{\gamma}^-_{k, \,(V_i,0)} \rangle = 1$.
This, together with Eq.~\eqref{eq:GammaQuench}, consequently implies:
\beq
   p_k^+(\varphi) = \sin^2 (\Delta \Theta_k) \,, \qquad
   p_k^-(\varphi) = \cos^2 (\Delta \Theta_k) \,.
   \label{eq:prob_quench}
\eeq

Having evaluated all the necessary quantities, namely Eq.~\eqref{eq:prob_quench}
for the probabilities and Eq.~\eqref{eq:HamDiag} for the energies, 
let us now compute both the stiffness, Eq.~\eqref{eq:stiff}, and the helicity modulus,
Eq.~\eqref{eq:helic}, for the various types of quenches.

\subsubsection{Quench from the insulator to the superfluid  ($V_i \neq 0$, $V_f = 0$)}

In the case $V_f = 0$, the dispersion relation for the Hamiltonian after the quench
is given by $\epsilon_k (0,\varphi) = 2 t \cos (k-\varphi)$, 
while the probabilities of the diagonal ensemble simplify to
$p_k^+(\varphi) = \sin^2 [\Theta_k(V_i,0)]$.

$\bullet$ In order to compute the stiffness, we apply directly Eq.~\eqref{eq:stiff}.
For this purpose it is useful to note that 
$\partial^2_\varphi \epsilon_k(0,\varphi) {}_{\vert_{\varphi \to 0}} = -2 t \cos(k)$.
Moreover we also observe that $\sin^2 (x) = [1-\cos(2x)]/2$ and that
$\cos [2 \Theta_k(V_i,0)] =  2 t \cos(k) / \epsilon_k (V_i,0)$.
Thus we obtain, in the continuum limit $L \to \infty$:
\beq
   \mathcal{D}_{\rm neq} = 
   \frac{V_i}{\pi} \left[ \mathbb{E} \bigg( \! \frac{-4 \, t^2}{V_i^2} \! \bigg) 
     - \mathbb{K} \bigg( \! \frac{-4 \, t^2}{V_i^2} \! \bigg) \right] \,,
   \label{eq_stiff_Q1}
\eeq
where $\mathbb{E}(m)$ and $\mathbb{K}(m)$ 
are the complete elliptic integrals of the second and of the first kind respectively, 
expressed in terms of the corresponding incomplete integrals given in Eq.~\eqref{eq:elliptic2}
and by
\beq
   \mathbb{K}(\phi \vert m) = \int_0^\phi (1 - m \sin^2 \theta)^{-1/2} {\rm d}\theta \,.
   \label{eq:elliptic1}
\eeq
The expression~\eqref{eq_stiff_Q1} tends to $\mathcal{D}_0$ for $V_i \to 0$,
and goes as $\mathcal{D}_{\rm neq} \sim t^2/V_i$ for $V_i \gg 1$ 
(see Fig.~\ref{exactquench}, upper left panel).

$\bullet$ Since in this specific case the probability $p_k^+(\varphi)$ 
does not depend on $\varphi$, from Eq.~\eqref{eq:helic} it is straightforward to see that
$\mathcal{Y}_{\rm neq} = \mathcal{D}_{\rm neq}$.

\subsubsection{Quench from the superfluid to the insulator ($V_i = 0$, $V_f \neq 0$)}

In the case $V_i = 0$, the dispersion relation after the quench is
given by $\epsilon_k (V_f,\varphi) = \sqrt{[2 t \cos (k-\varphi)]^2 + V_f^2}$,
while the probabilities of the diagonal ensemble are 
$p_k^+(\varphi) = \sin^2 [\Theta_k(V_f,\varphi)]$.

$\bullet$ Also in this case it is better to compute the stiffness directly from Eq.~\eqref{eq:stiff}.
After adopting the same type of substitutions leading to Eq.~\eqref{eq_stiff_Q1},
this produces the following expression in the thermodynamic limit:
\beq
   \mathcal{D}_{\rm neq} = \mathcal{D}_0 \Bigg[ \frac{1}{2} - \frac{1}{8t} \frac{V_f^2}{\sqrt{4 t^2 + V_f^2}}
     \log \bigg( \frac{8 t^2 + V_f^2 + 4 t \sqrt{4 t^2 + V_f^2}}{V_f^2} \bigg) \Bigg] \,.
   \label{eq_stiff_Q2}
\eeq
While, for $V_f \to 0$, Eq.~\eqref{eq_stiff_Q2} tends to $\mathcal{D}_0 / 2$,
for large $V_f$ values it goes to zero as $\mathcal{D}_{\rm neq} = 4 \mathcal{D}_0 t^2 /3 V_f^2$ 
(see Fig.~\ref{exactquench}, upper right panel).

$\bullet$ For the evaluation of the helicity modulus, in this specific case we
can easily use Eq.~\eqref{eq:helic_fin}, since, as in equilibrium, the dependence of both
the dispersion relation $\epsilon_k (V_f,\varphi)$ and the probabilities $p_k^\pm(\varphi)$
on the flux $\varphi$ always come as $k-\varphi$.
Taking the derivative with respect to the momentum of the energies 
$\varepsilon_k^\pm (V_f,0) = \pm \sqrt{(2t \cos k)^2 + V_f^2}$
and evaluating it at the extrema we have:
\beq
   \frac{\partial \varepsilon_k^\pm (V_f,0)}{\partial k} \bigg\vert_{-\pi/2}^{\pi/2} =
   \mp \frac{2 t^2 \sin(\pm \pi)}{V_f} \,.
\eeq
Therefore it is evident that, for any $V_f \neq 0$, the helicity modulus is rigorously zero.

\subsubsection{Quench within the insulating phase ($V_i \neq 0$, $V_f \neq 0$)}

$\bullet$ In the generic case of both initial and final values of $V$ different from zero, 
we proceed by substituting the expressions~\eqref{eq:HamDiag} and~\eqref{eq:prob_quench} 
into Eq.~\eqref{eq:stiff}.
For this purpose, it is useful to note that $p_k^+(\varphi) - p_k^-(\varphi) = 2 p_k^+(\varphi) -1
= - \cos(2 \, \Delta \Theta_k) = -\cos [2 \, \Theta_k(V_f,\varphi) ] \, \cos [2 \, \Theta_k(V_i,0) ]
- \sin [2 \, \Theta_k(V_f,\varphi) ] \, \sin [2 \, \Theta_k(V_i,0) ]$.
Hence, using the fact that $\cos [2 \, \Theta_k] =  2 t \cos(k) / \epsilon_k$
and that $\sin [2 \, \Theta_k] =  V / \epsilon_k$, 
we easily obtain a formula for the stiffness that is suitable for numerical calculations:
\beq
   \mathcal{D}_{\rm neq} = \frac{1}{2 \pi} \int_{-\pi/2}^{\pi/2} 
    \left\{ \frac{(2 t \cos k)^2 + V_i V_f}{\epsilon_k(V_i,0) \, \epsilon_k(V_f,0)} \right\} \;
   \bigg[ \partial_k \frac{2 t^2 \sin (2k)}{\epsilon_k(V_f,0)} \bigg]  {\rm d}k \,.
\eeq
The term in curl brackets, hereafter called $\alpha$, comes from the populations 
$p_k^+(\varphi) - p_k^-(\varphi)$.

While for $\Delta V \equiv V_i - V_f = 0$ we have $\alpha=1$ 
(hence $\mathcal{D} = 0$ for any $V > 0$ in equilibrium),
for a small quench $\vert \Delta V \vert \gtrsim 0$ we find:
\beq
   \alpha \simeq 1 -\frac{2 \, t^2 \cos^2 k}{[\epsilon_k(V_f,0)]^4} \Delta V^2 + \ldots
\eeq
Therefore inserting it into the previous expression and integrating, we finally arrive to
\beq
   \mathcal{D}_{\rm neq} \simeq \mathcal{D}_0 \, \Delta V^2 \, \mathcal{C}(V_f)
\eeq
with
\beq
   \mathcal{C}(V) 
   = \left[ (8 t^4 + 22 t^2 V^2 + 3 V^4) \;\, \mathbb{E} \! \left( \! \frac{4 t^2}{4 t^2+V^2} \! \right) 
   - V^2 (16 t^2 + 3V^2) \; \mathbb{K} \! \left( \! \frac{4 t^2}{4 t^2 +V^2} \! \right) \right]
   \bigg/ \left[ 15 t \, V^2 (4 t^2 + V^2)^{3/2} \right] \,.
\eeq
The function $\mathcal{C}(V)$ is positive for $V \lesssim 0.7 \, t$
and diverges for $V \to 0$, while it becomes negative otherwise,
with a minimum at $V \sim t$.
Hence out of equilibrium the stiffness can be negative (see Fig.~\ref{exactquench}).

$\bullet$ Equation~\eqref{eq:helic_fin} for the helicity modulus 
cannot be used in this context, since from Eq.~\eqref{eq:prob_quench} and
the expression for $\Delta \Theta_k$ in Eq.~\eqref{eq:GammaQuench} it is evident that
the dependence on the angle $\varphi$ of the probabilities $p_k^\pm$ is not simply through $k-\varphi$.
We then proceed using Eq.~\eqref{eq:helic}, and in particular focusing on
the last term $\beta = \mathcal{Y}_{\rm neq} - \mathcal{D}_{\rm neq}$.
We first observe that $\partial_\varphi (p_k^+ - p_k^-)
= - \{ \partial_\varphi \cos [2 \, \Theta_k(V_f) ] \} \, \cos [2 \, \Theta_k(V_i) ]
  - \{ \partial_\varphi \sin [2 \, \Theta_k(V_f) ] \} \, \sin [2 \, \Theta_k(V_i) ]$.
Then, after simple algebraic manipulations, we find the following expression:
\beq
   \beta = - \mathcal{D}_0 \; \Delta V \int_{-\pi/2}^{\pi/2}
   \frac{t^3 V_f \sin^2(2k)}{\epsilon_k(V_i) \; [\epsilon_k(V_f)]^4 }  {\rm d}k \,.
\eeq
This clearly confirms that in equilibrium $\beta = 0$
and therefore $\mathcal{D} = \mathcal{Y}$.
Now, if we suppose to perform a small quench, $\vert \Delta V \vert \gtrsim 0$,
we can approximate the integrand denominator with $[\epsilon_k(V_f)]^5$
and thus finally arrive at the expression:
\beq
   \mathcal{Y}_{\rm neq} \approx \mathcal{D}_{\rm neq} + \mathcal{D}_0 \; \Delta V \; \tilde{\mathcal{C}}(V_f)
   = \mathcal{D}_{\rm 0} \left\{ \Delta V^2 \, \mathcal{C}(V_f) + \Delta V \; \tilde{\mathcal{C}}(V_f) \right\}\,,
\eeq
where the term $\tilde{\mathcal{C}}$ is given by
\beq
   \tilde{\mathcal{C}}(V) = 
   \frac{1}{3t} \left\{ \mathbb{K} \! \left( \! -\frac{4 t^2}{V^2} \! \right) 
   - \left[ \frac{2t^2 + V^2}{4t^2 + V^2} \right] \; 
   \mathbb{E} \! \left( \! -\frac{4 t^2}{V^2} \! \right) \right\} \,.
\eeq
This function is always negative and monotonic increasing from  $\mathcal{C} \to - \infty$
for $V \to 0$, to zero in the limit $V \to \infty$ 
(see Fig.~\ref{exactquench}, lower panel).

\section{Non-integrable model of hard-core bosons}
\label{sec:Model2_IB}

We analyse now a modified version of Model II for interacting hard-core bosons, 
in which we add an integrability-breaking perturbation. The integrable Hamiltonian is given by:
\beq
   \hat{\Ham}(\varphi) = 
   t \sum_{j=1}^L \left( e^{i \varphi} \hat{b}^\dagger_{j} \hat{b}_{j+1} 
   + e^{-i\varphi} \hat{b}^\dagger_{j+1} \hat{b}_j \right)
   + V \sum_{j=1}^L \hat{n}_j \hat{n}_{j+1} \,. 
   \label{eq:model2_IB}
\eeq
There are several ways to break the integrability in this system.
Here we choose to perturb it by adding a random on-site chemical potential of the form
\beq
   \hat{\cal K} = \lambda \sum_{j=1}^L h_j \, \hat{n}_j \,,
   \label{eq:IB_perturb}
\eeq
where $h_j$ are random real variables, $h_j \in [-1,+1]$, and $\lambda$ denotes the
strength of the perturbation.
The global non-integrable Hamiltonian we are going to study here is thus
$\hat{\Ham}_{\rm IB}(\varphi) = \hat{\Ham}(\varphi) + \hat{\cal K}$.
According to the standard theory of quantum chaotic systems, we believe that 
the specific choice of $\hat{\cal K}$ is not qualitatively 
relevant and does not affect our conclusions.

In the specific, we provide numerical evidence in support of our conjecture 
that, when integrability is broken, the average stiffness after a quench should 
no longer be negative.
The reason resides in the fact that, in such cases, thermalisation generally set up
and therefore the system is expected to locally behave as if it were 
in a canonical thermal ensemble at an effective
temperature determined by the energy after the quench.

Hereafter in this Section we discuss only the case of a quench towards a superfluid phase
and take $t$ as the energy unit.
In the specific we set $V_f = 0.5$, corresponding to the situation of left panel 
of Fig.~2 in the main text, where already for small sizes we were able to detect
a dramatic decrease of ${\cal D}_{\rm neq}$ towards very large negative values.
Due to numerical limitations of our finite-size scaling, it is more difficult to 
highlight the effects of an integrability-breaking perturbation in the cases 
where the stiffness for the integrable model is already close to zero and negative
(such as for a quench towards the insulator, as in the right panel of Fig.~2).

\subsection{Level spacing statistics}
\label{sec:LSS}

Let us first discuss the onset of quantum chaos for the Hamiltonian $\hat{\Ham}_{\rm IB}$.

The distinctive signature of a non-integrable system is the tendency to
favour avoided level crossings, due to the lack of non-trivial constants
of motions other than the energy.
On the contrary, integrable systems have levels which tend to cluster,
and cross when some parameter in the Hamiltonian is varied.
A quantitative indicator of such features is the level spacing statistics (LSS)
of the energy differences between adjacent levels 
in the Hamiltonian spectrum~\cite{Haake}.
This indicator is constructed such that $P(s)$ represents 
the probability to find the quantity $s_n \equiv E_{n+1} - E_n$ 
(normalised to the average level spacing) in the interval $[s,s + ds]$. 

A typical integrable system obeys a Poissonian (P) statistics: 
\beq
   P_{\rm P} (s) = e^{-s} \,.
   \label{eq:LSS-P}
\eeq
Vice-versa, the statistics of a non-integrable system is generally well described 
by the random-matrix theory, leading to a Wigner-Dyson (WD) distribution~\cite{Haake}.
In particular, the above mentioned level repulsion emerges in the fact that 
$P_{\rm WD} \stackrel{s \to 0}{\longrightarrow} s^\gamma$,
where $\gamma$ depends on the symmetries of the model.
In our specific case, Eq.~\eqref{eq:model2_IB} is invariant under time-reversal, 
therefore its LSS is given by a Gaussian orthogonal ensemble:
\beq
   P_{\rm WD}(s) = \frac{\pi s}{2} e^{- \pi s^2/4} \,.
   \label{eq:LSS-WD}
\eeq

We found that the system described by Eq.~\eqref{eq:model2_IB},
together with the integrability-breaking perturbation~\eqref{eq:IB_perturb},
undergoes a transition from Poissonian to Wigner-Dyson LSS upon
increasing the perturbation strength $\lambda$ (see Fig.~\ref{fig:LSS} below). 
This transition is better quantified by means of the level spacing indicator (LSI):
\beq
   \eta = \frac{\int [P(s) - P_{\rm WD}(s)] \, {\rm d}s}{\int [P_{\rm P}(s) - P_{\rm WD}(s)] \, {\rm d}s},
\eeq
so that $\eta = 0$ for systems obeying a Wigner-Dyson distribution of spacings,
while $\eta = 1$ if the distribution is Poissonian.

\begin{figure}[b]
  \includegraphics[width=0.7\columnwidth]{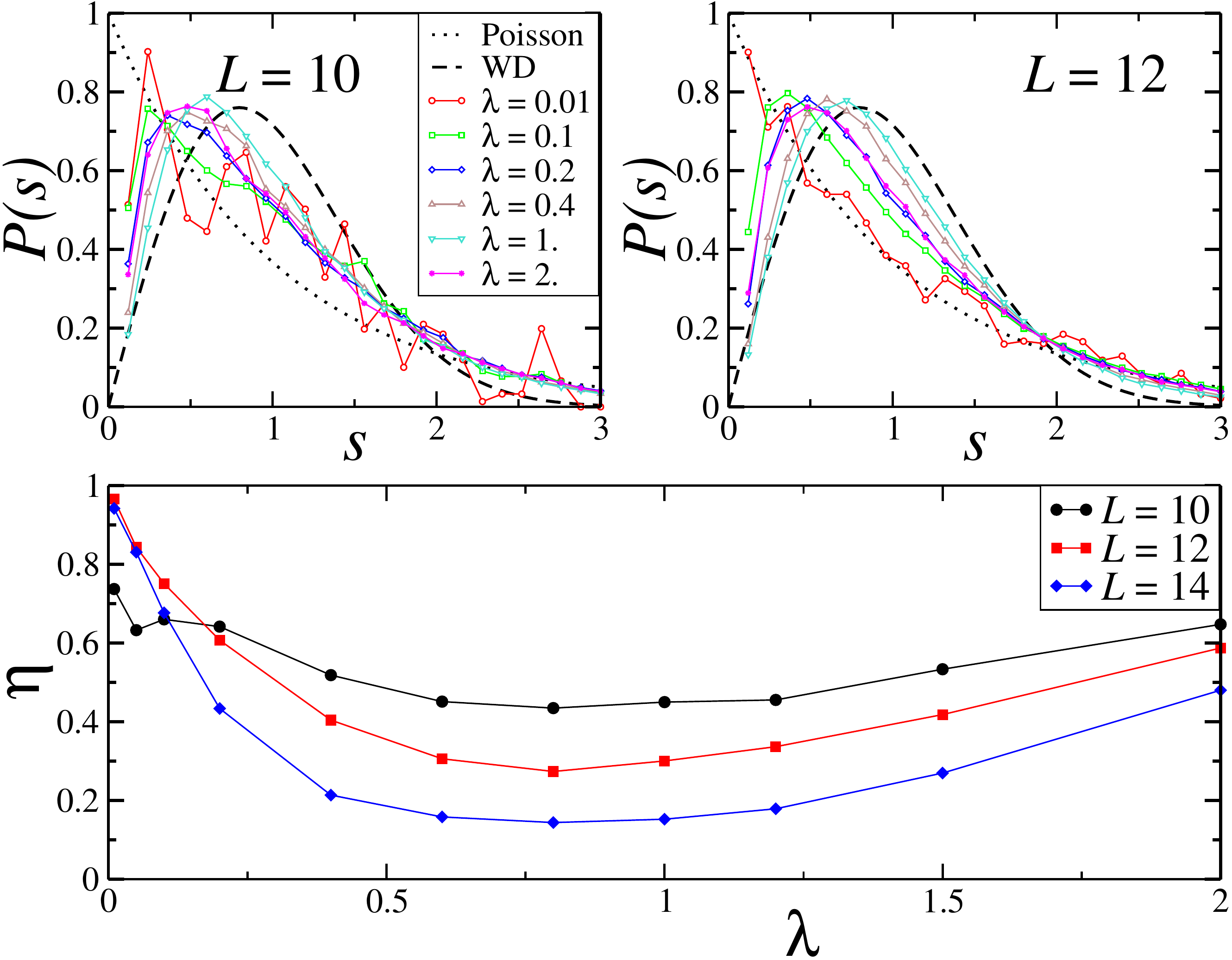}
  \caption{(colour online) Upper panels: level spacing statistics
    for the model in Eq.~\eqref{eq:model2_IB}, together with the 
    integrability-breaking perturbation~\eqref{eq:IB_perturb}.
    Here we consider $V = 0.5$ and work in units of $t=1$.
    The various curves are for different values of the perturbation strength $\lambda$.
    The dashed black line indicates the Wigner-Dyson behaviour~\eqref{eq:LSS-WD},
    while the dotted black line is the Poissonian~\eqref{eq:LSS-P}.
    Lower panels: level spacing indicator as a function of $\lambda$ and for various
    system sizes. Note the increasing sharpness and convergence to the WD
    for intermediate $\lambda$ values, when $L$ is increased.
    Averages have been performed over $10^{4}$, $10^3$, $10^2$ disorder realisations,
    for $L=10, 12, 14$ respectively.}
  \label{fig:LSS}
\end{figure}

Since the specific form of Eq.~\eqref{eq:LSS-WD} 
corresponds to the case in which only a time-reversal symmetry is present, 
in order to remove any other unwanted symmetry in our simulations for the level statistics 
we considered open boundary conditions, fixed the sector of zero magnetisation, 
and added a very small magnetic field on the first site of the chain.
In our counting for $P(s)$, we also cut the lowest and the largest energy levels, 
since integrability breaking typically manifests only in the middle of
the spectrum, where the levels repel.
The net effect of all these expedients should become negligible in the thermodynamic limit,
while we expect they quantitatively affect our finite-size results.

The results displayed in Fig.~\ref{fig:LSS} clearly show that, 
by fixing the system size $L$, if $\lambda$ is progressively
increased, the LSS goes from a nearly Poissonian shape (for $\lambda \approx 0$)
to a form closely resembling the WD (for $\lambda \approx 0.8$).
For $\lambda \gtrsim 1$ it returns closer to Poisson, since in that case
the system approaches a trivial integrable model ($\lambda \gg 1$).
This non-monotonic behaviour shows up more evidently in the LSI, 
which displays a minimum around $\lambda \approx 0.8$.
We stress that the effect of the integrability breaking is more evident for large $L$, 
where the convergence to the WD distribution with $\lambda$ improves
({\it e.g.}, note that for $L=10$, $P(s)$ is still quite far from
WD at $\lambda \approx 0.8$---see the upper left panel of Fig.~\ref{fig:LSS}).

\subsection{Stiffness after a quench}
\label{sec:Stiff_IB}

We now address the behaviour of the averaged stiffness ${\cal D}_{\rm neq}$
after a sudden quench in the interaction $V$.
As noted in the main text, in out-of-equilibrium conditions ${\cal D}_{\rm neq}$
can assume negative values.
However we verified that, by switching on gradually an integrability-breaking term 
in the Hamiltonian, such a situation tends to disappear and the stiffness
becomes positive for any type of quench.
Of course, due to the finiteness of the systems we are considering, 
by increasing $\lambda$ we observe a crossover from negative to positive values 
of the stiffness.

Figure~\ref{fig:Stiff_IB} displays a typical situation where, in absence of 
non-integrable perturbation, when quenching from a ground state deep in the insulator
towards the superfluid, model II exhibits large negative values of ${\cal D}_{\rm neq}$
(black curve).
However, as long as we increase the strength $\lambda$ of such perturbation,
the values are in modulus drastically reduced and the stiffness eventually becomes always 
positive for $\lambda \approx 0.8$, that is when $\eta$ gets closer to zero and the LSS approaches WD.
To make more precise connections one would have to go to much larger sizes,
where the smooth crossover should sharpen.

\begin{figure}[!h]
  \includegraphics[width=0.7\columnwidth]{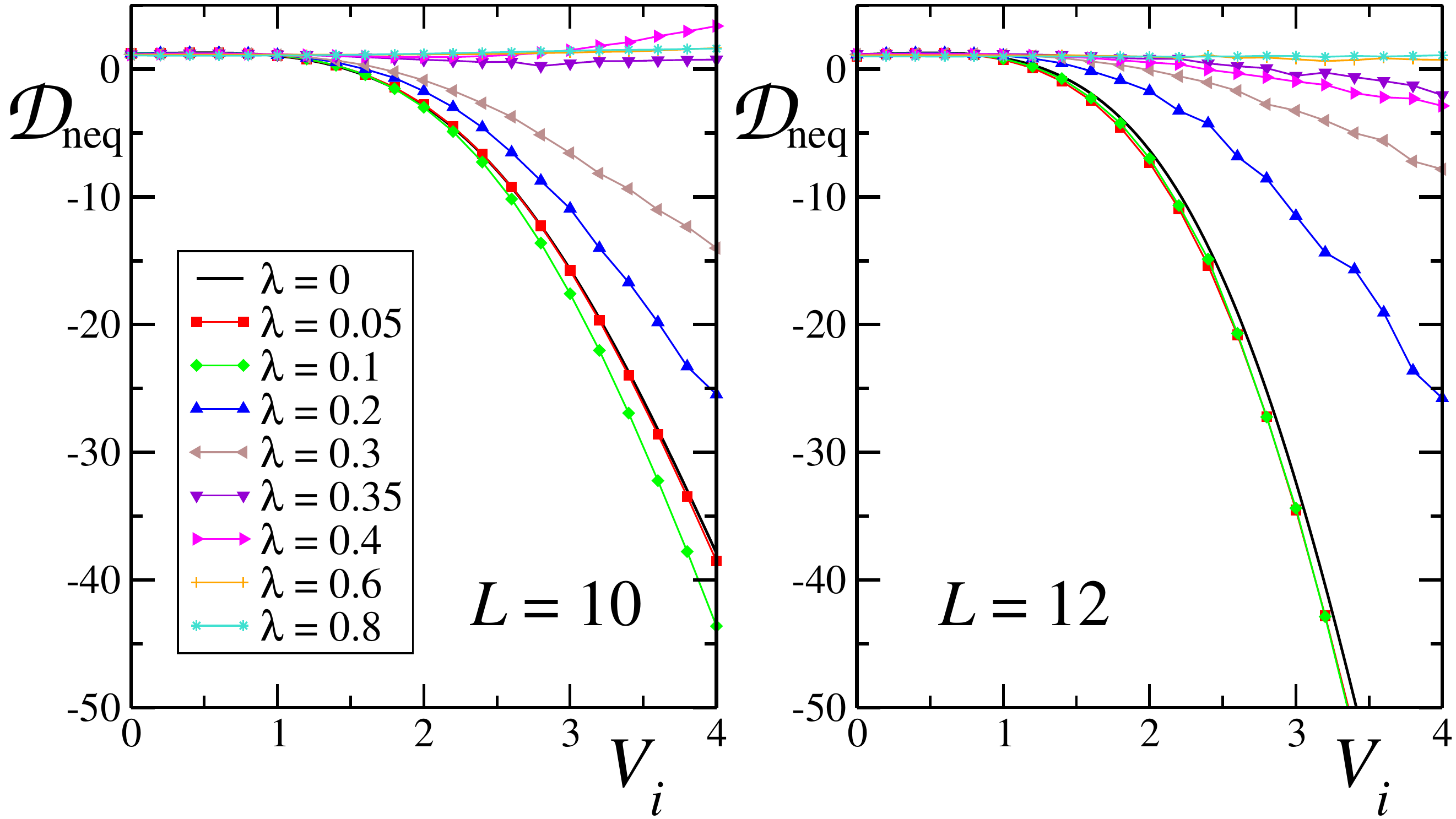}
  \caption{(colour online) Average stiffness ${\cal D}_{\rm neq}$ for $\hat{\Ham}_{\rm IB}$
    as described by Eq.~\eqref{eq:model2_IB}, plus the integrability-breaking 
    perturbation of Eq.~\eqref{eq:IB_perturb} of intensity $\lambda$.
    The stiffness has been computed for a quench towards the superfluid
    phase with $V_f = 0.5$, starting from various initial ground states 
    of the same Hamiltonian with interaction strength $V_i$.
    Data have been averaged over $5 \times 10^3$ and $5 \times 10^2$ disorder
    realisations, for $L=10$ (left panel) and $L=12$ (right panel) respectively.}
  \label{fig:Stiff_IB}
\end{figure}

\end{widetext}


\begin{thebibliography}{100}
\bibitem{note_0}
  For integrable systems, diagonal ensemble averages of local quantities, 
  such as the current, should coincide with averages performed on a GGE.
\bibitem{note_2}
  This condition is equivalent to have $V^\star = 2 \cos (k \pi / K) $ with $k = 0, \ldots, K-1$.
\end{thebibliography}
\end{document}